\documentclass[twocolumn]{aastex701}

\usepackage{amsmath} 
\usepackage{hyperref}
\usepackage{booktabs}

\begin{document}

\title{Modeling Multi-band Active Galactic Nuclei Lightcurves Using a Self-supervised Attentive Neural Process Network}

\author[orcid=0009-0005-8032-0200]{Tzuken Shen}
\affil{School of Engineering and Applied Sciences, Harvard University, Cambridge, MA 02138, USA.}
\affil{Center for Astrophysics, Harvard \& Smithsonian, 60 Garden Street, Cambridge, MA 02138, USA}
\email[show]{kenshen@g.harvard.edu}  

\author[0000-0003-3106-8182]{Chengcheng Xin}
\affil{Harvard University, Black Hole Initiative, 20 Garden Street, Cambridge, MA 02138, USA}
\affil{Center for Astrophysics, Harvard \& Smithsonian, 60 Garden Street, Cambridge, MA 02138, USA}
\email{chengcheng.xin@cfa.harvard.edu}

\author[0000-0001-8723-6136]{Joshua~Fagin}
\affiliation{The Graduate Center of the City University of New York, 365 Fifth Avenue, New York, NY 10016, USA}
\affiliation{Department of Astrophysics, American Museum of Natural History, Central Park West and 79th Street, NY 10024-5192, USA}
\affiliation{Department of Physics and Astronomy, Lehman College of the CUNY, Bronx, NY 10468, USA}
\email{}

\author[0000-0002-5814-4061]{Ashley Villar}
\affil{Center for Astrophysics, Harvard \& Smithsonian, 60 Garden Street, Cambridge, MA 02138, USA}
\email{ashleyvillar@cfa.harvard.edu}

\author[0000-0002-3168-0139]{Matthew J. Graham}
\affiliation{Department of Physics, Maths and Astronomy, California Institute of Technology, 1200 E California Blvd, Pasadena, CA 91125, USA}
\email{}

\author[0000-0002-5956-851X]{K.E. Saavik Ford}
\affiliation{Department of Science, BMCC, City University of New York, New York, NY 10007, USA}
\affiliation{Department of Astrophysics, American Museum of Natural History, New York, NY 10024, USA}
\affiliation{Graduate Center, City University of New York, 365 5th Avenue, New York, NY 10016, USA}
\email{}

\begin{abstract}

With the the Legacy Survey of Space and Time (LSST), observations
of tens of millions of quasars are expected. We present a machine-learning
framework based on an attentive neural process (ANP) that models stochastic
quasar variability and jointly performs two
tasks: (i) probabilistic reconstruction of quasar light curves in all six
LSST bands, and (ii) inference of physical parameters of the AGN system ---
black hole mass, Eddington ratio, and intrinsic variability timescale --- with joint, full-covariance posteriors. The model
processes all six bands within a single masked attention mechanism, so
that observations in any band constrain the shared driving process, and
natively handles irregularly and asynchronously sampled light curves with
realistic heteroscedastic noise. We train and evaluate on full ten-year
light curves simulated with a relativistic accretion-disk reprocessing
model, observed through the fielded Rubin baseline cadence with per-visit
photometric noise. The light-curve model is trained with self-supervised,
learning from the sparse noisy observations alone and never accessing the
underlying noiseless light curves. On a held-out test set, we show that our ANP model reach comparable performances to that of the near-optimal baseline Gaussian process model, while
exceeding it in power spectrum reconstruction and damped-random walk timescale estimation. The
compact learned latent representation supports future downstream applications,
and the fully observation-driven training objective enables direct
fine-tuning on real LSST light curves.
\end{abstract}

\keywords{\uat{Active galactic nuclei}{16} --- \uat{Supermassive black holes}{1663} --- \uat{Time domain astronomy}{2109}}


 \section{Introduction} 

Quasars, powered by accretion onto supermassive black holes (SMBHs) at the centers of galaxies, are among the most luminous sources in the Universe \citep{LyndenBell1969,Netzer2015}. They exhibit temporal variability across the electromagnetic spectrum \citep[e.g.][]{Ulrich1997,Giveon1999,McHardy2006,Soldi2008,Edelson2015}, with the UV and optical brightness of quasars varying by at least 10\% on weeks to years timescales \citep{VandenBerk2004,Wilhite2005}. The Legacy Survey of Space and Time (LSST) by the Vera C. Rubin Observatory has officially begun its 10 year observations; it is expected to observe $\sim 10^7-10^8$ quasars across a wide range of masses ($\sim 10^5-10^{10} {\rm M_{\odot}}$) and up to high redshifts ($z\sim 6$) \citep{LSSTScienceCollaboration2009,Xin2021,Viitanen2026AGILE}. %The LSST Deep-Drilling Field (DDF) survey will deliver a large sample of quasars %down to very faint magnitude limits of $i$-mag = 27 
% with long baseline of $10$ years and and high-frequency sampling with effective cadence of $\sim 1$ day in any filter \citep{LSSTScienceCollaboration2009}. This allows us to probe both short-term and long-term quasar variability. 

The smooth and non-periodic temporal spectrum of quasars suggests that their variability is stochastic in nature, with correlated, frequency-dependent continuum and emission-line variations \citep{Kelly2009,MacLeod2010,Wilhite2005,Pereyra2006}. 
The UV and optical flux variations are believed to be either directly driven by the accretion onto the SMBH via a geometrically thin and optically thick accretion disk \citep{ShakuraSunyaev1973,Pereyra2006,Liu2008} or by reprocessing the X-ray radiation from the hot corona \citep{BlandfordMcKee1982,Wanders1997,Cackett2021,Wang2023}. These variations provide a time-domain probe of spatial scales that are generally too small to resolve directly.

% In a sample of more than 25,000 quasars, \cite{VandenBerk2004} found a pronounced anti-correlation between variability amplitude and rest-frame wavelength, together with an anti-correlation between variability amplitude and luminosity. 
% 
Using a large sample of SDSS Stripe 82 quasars, \citet{MacLeod2010} have confirmed these trends and further shown that quasar variability depends on black hole mass, Eddington ratio, radio loudness, rest-frame wavelength, among other source properties. They modeled the optical variability of active galactic nuclei (AGNs\footnote{In the remainder of this paper, we use the broader term active galactic nucleus (AGN) instead of quasar, because our analysis considers a wide range of active galaxies expected to be observable with LSST.}) using the damped random walk (DRW) process, a framework originally proposed by \citet{Kelly2009}. 
The DRW is a stationary stochastic process characterized by a damping timescale, ($\tau_{\rm DRW}$), and a long-term asymptotic variability amplitude, (SF$_{\infty}$). Its power spectral density has the form of a broken power law, which is approximately flat at low frequencies ($\gg 1/\tau_{\rm DRW}$) below the characteristic break and drops as ($f^{-2}$) at higher frequencies. Equivalently, the structure function,  corresponding to the long-term DRW variability amplitude, increases with time lag and approaches SF$_{\infty}$ at lags much longer than $\tau_{\rm DRW}$. On these long timescales, pairs of measurements become effectively uncorrelated, and the variability amplitude no longer increases with temporal separation. 

% %%%

Despite its success, the DRW model is incomplete in describing the full diversity of AGN light curves. For dxample, high-cadence observations in Kepler have revealed a steeper power-spectral density at short timescales \citep{Mushotzky2011}. 
% \cx{Are we also limited on the longer timescale due to baseline? } 
More flexible stochastic models have since then  been developed, e.g. the damped harmonic oscillator model \citep{Yu2022}, and such models seem to better capture population-level AGN behavior \citep{yu2025examining}.

Recent studies of AGN variability have increasingly incorporated
simulation-based machine learning techniques. In this approach, physically
motivated models of X-ray variability and accretion-disk reprocessing are
used to generate large sets of simulated AGN light curves, which are then
used to train neural-network models. For example, \citet{Fagin_2025} used a
latent stochastic differential equation (SDE) framework to model AGN
light-curve variability, reconstruct missing seasonal gaps, and infer
accretion-disk physical parameters. While this approach provides a
physically motivated and self-consistent framework, it also requires a
complex model architecture and substantial computational resources for
training. A complementary line of work builds on the neural process (NP)
family, which combines the flexibility of deep learning with the amortized
probabilistic inference of Gaussian processes.
\citet{cvorovic2021conditional} first applied conditional neural processes
to AGN light curves, demonstrating flexible non-parametric interpolation of
survey data, though without inferring physical parameters or assessing the
calibration of the predictive uncertainties.
\citet{park2021inferringblackholeproperties} extended the family to black
hole parameter inference with an attentive neural process (ANP); their model
is compact, but its architecture assumes synchronous multi-band
observations, limiting its direct applicability to realistic LSST light
curves, which are irregularly sampled and asynchronous across bands. Most
recently, \citet{raju2026metalearning} combined attentive latent neural
processes with self-organizing-map clustering and mixture density networks
to recover accretion-disk transfer functions and SMBH parameters; their
framework treats each photometric band independently, trains on noiseless
synthetic light curves with short, burst-like cadences, and validates
parameter recovery at the level of population distributions rather than
individual objects.

In this work, we propose an extended ANP model based on the framework of
\citet{park2021inferringblackholeproperties}. Our model retains the compact
latent-representation structure of the ANP while adapting it to realistic
LSST-like observations. Compared with the latent-SDE framework, the ANP
provides a simpler and more modular architecture for inference from
observed light curves, and its compact learned latent representation can be
used not only for light-curve reconstruction and AGN parameter inference
but also for further downstream tasks; for example, the latent space allows
simple clustering, potentially identifying subclasses of AGN. Compared with
the existing NP-based approaches, our model advances the family along four
axes: it processes all six LSST bands jointly within a single masked
attention mechanism, so that observations in any band constrain the shared
driving process; it is trained and evaluated on visit lists drawn from the
fielded Rubin observing strategy with per-visit photometric noise, rather
than on synthetic cadences or noiseless targets; its light-curve model is
trained in a fully self-supervised manner, learning exclusively from the
sparse, noisy observations themselves; and it produces joint,
full-covariance posteriors over four physical parameters whose calibration
we verify per object, through the empirical coverage of individual credible
intervals.

The self-supervised character of the reconstruction training deserves
emphasis, as it determines the framework's applicability beyond
simulations. The model is never shown the underlying noiseless light
curves. Instead, we design a tokenization pipeline that handles irregularly and
asynchronously sampled multi-band observations
(\S~\ref{subsec:data_tokenization}), and the loss function is
evaluated only against held-out \emph{observed} photometry, in
contrast to approaches that train against noiseless simulated targets.
Only the parameter-inference decoder is trained in a supervised approach, using the known
simulation parameters. Because real LSST light curves provide exactly the
same information the reconstruction model trains on --- noisy, irregular
photometry with error bars, and no ground truth --- the framework can be
fine-tuned on the millions of real AGN light curves LSST will deliver
without any change to the training objective.

We start our analysis by simulating quasar lightcurves that will be observable by LSST in its 10-year survey, in \S~\ref{sec:LC_sims}, starting from generating an accretion disk model that reprocesses X-ray emissions into UV/optical variability as well as simulating LSST cadences and noise. We then describe the neural network framework used to model the AGN parameters in \S~\ref{sec:modeling}, where we describe tokenization pipeline applied on the simulated multi-band LSST lightcurves and our attentive neural processing model. We present our results of the performance of the neural process modeling on inferring AGN variability parameters and AGN physical parameters. We find high effectiveness on measuring a wide range of DRW timescales and amplitudes, as well as predicting the BH mass, Eddington fraction and redshift. %With said parameters, we reconstruct multi-band LSST lightcurves to the negative log-likelihood with median of $-2.47 \pm 0.810$ over the testing light curves. 
We additionally find that the trained model is highly effective in reconstructing unseen multi-band LSST lightcurves in our test set.
Finally, we discuss future application of this work in \S~\ref{sec:discussion}.

\section{Light-curve Simulation} \label{sec:LC_sims}

We follow the basic approach of \citep{Fagin_2025} to generate simulated AGN light curves as they would be observed by LSST with slight modification. These simulated light curves are used to train and evaluate our ANP model.

The simulated AGN light curves are assumed to arise from geometrically thin, optically thick accretion disks irradiated by a central ``lamp-post'' X-ray source, commonly interpreted as a compact hot corona near the center of the quasar \citep{Chan_2025}. Variability in this central X-ray source is reprocessed by the accretion disk. Emission at different wavelengths originates predominantly from unique disk radii; in turn, different observational bands respond to the driving X-ray variability with different time delays and amplitudes. This produces a wavelength-dependent lag spectrum, $\tau_\lambda$, and flux spectrum, $f_\lambda$. The reprocessing can be described by a wavelength-dependent transfer function. The light curve in each photometric band is derived by convolving the driving X-ray variability with the corresponding transfer function. To simulate realistic observations, we apply realistic LSST observing cadence and properties to the reprocessed light curves. The details of each component of the simulation are described below.

\subsection{X-ray Variability Simulation}
A common model for describing the central X-ray variability of an AGN is the DRW process, also known as the Ornstein--Uhlenbeck process \citep{Kelly2009}. This stochastic process can be written as
\begin{align}\label{eq:DRW}
   dX(t) = -\frac{1}{\tau}X(t)\,dt + \sigma\,dW(t) + b\,dt ,
\end{align}
where $dW(t)$ is the increment of a Wiener process over some increment of time $dt$. Equivalently, in a time-discretized form that we use in this work, $dW(t)$ can be written as $\sqrt{dt}\,\epsilon(t)$, with $\epsilon(t)\sim N(0,1)$. Here, $\tau$ is the characteristic damping timescale, $b$ determines the long-term mean of the variability signal through $\bar{X}=b\tau$, and $\sigma$ controls the stochastic variability amplitude.

\subsection{Transfer Function Simulation}\label{sec:Transfer_func}

We model the reprocessing of photons by the accretion disk using reverberation mapping in the lamp-post framework \citep{Chan_2025}. Again, in this picture, fluctuations in the central X-ray source heat the disk and induce delayed, wavelength-dependent variations in the disk emission.

Starting from the standard thin-disk temperature profile, the disk temperature can be expanded to first order around its mean value as a function of radius $R$ and time $t$:
\begin{align}
    T(R,t)
    &\approx \bar{T}(R)
    \left[
    1+\frac{T_{\mathrm{lamp}}^4(R)}{4\bar{T}^4(R)}
    X\bigl(t-\tau\bigr)
    \right] \nonumber\\
    &= \bar{T}(R)+\Delta T(R)\,X\bigl(t-\tau\bigr).
\end{align}
Here, $X(t-\tau)$ is the driving variability signal, modeled by the DRW process (Eqn.~\ref{eq:DRW}) with a time delay $\tau$. The quantity $T_{\mathrm{lamp}}$ describes the contribution to the disk temperature from irradiation by the central lamp-post source, while $\bar{T}$ is the mean effective disk temperature. The mean temperature includes both viscous heating and external irradiation:
\begin{align}
    \bar{T}^4(R) \;&=\; T_{\mathrm{visc}}^4(R) \;+\; T_{\mathrm{lamp}}^4(R)
\;\nonumber\\&=\; \left(\frac{3\,G\,M_{\mathrm{bh}}\,\dot{M}_{\mathrm{in}}}
{8\,\pi\,\sigma_{\mathrm{SB}}\,R^3}\right)\ \times\nonumber\\
&\left[
\left(\frac{R_{\mathrm{in}}}{R}\right)^{3-\beta}
\left(1 - \sqrt{\frac{R_{\mathrm{in}}}{R}}\,\right)
\;+\;
\frac{4\,\eta_{X}\epsilon\,(1-A)\,R_{\mathrm{in}}^{3}}
{3\,\bigl(R^2 + H^2\bigr)^{3/2}}
\right]
\end{align}

Then, assuming blackbody radiation with a color-correction factor $f_{\mathrm{col}}$, the specific intensity at frequency $\nu$ is
\begin{align}
    B_{\nu}(T,t)
    &= \frac{2\,h\,\nu^3}{f_{\mathrm{col}}^3 c^2}
    \frac{1}{\exp\!\left(\frac{h\nu}{f_{\mathrm{col}} k_B T}\right)-1}
    \nonumber\\
    &\approx \overline{B_{\nu}}(R)
    + \Delta B_{\nu}(R)\,X\bigl(t-\tau\bigr).
\end{align}
Integrating $B_{\nu}$ over the solid angle gives the observed flux spectrum of the reprocessed emission. The time delay $\tau$ arises from the photon travel time between the central X-ray source and a given disk radius. Since each radius has a characteristic temperature, and therefore contributes preferentially to emission at particular wavelengths, the disk naturally produces wavelength-dependent time delays.

\begin{figure}[t]
    \centering
    \includegraphics[width=1.0\linewidth]{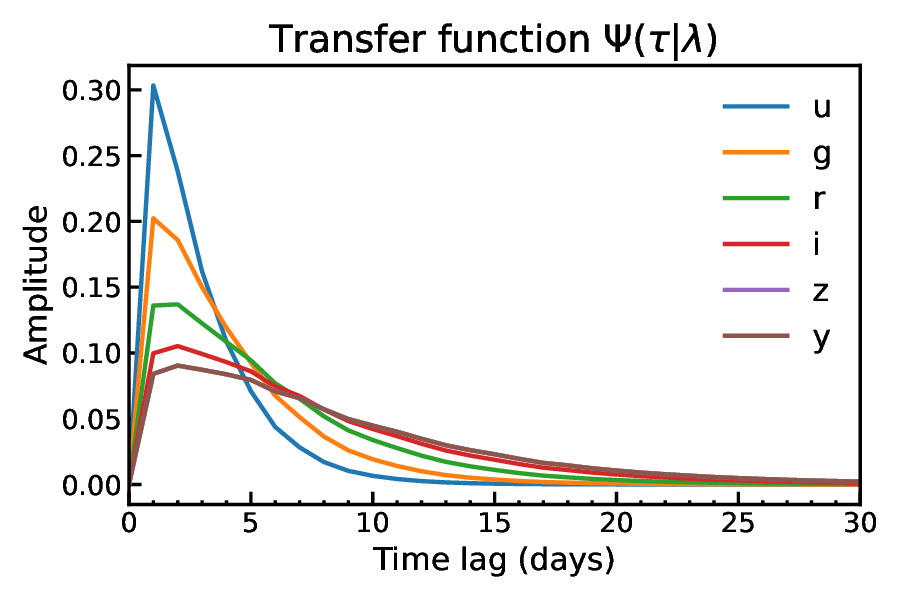}
    \caption{Example wavelength-dependent transfer functions, $\Psi_\nu(\tau|\lambda)$, for the six LSST bands. The black hole parameters in this example are: $\log(M_{bh})=8.0$,~$a=0$, ~$\log(\lambda_{edd})=-2.3$, ~$H=12$, ~$\theta_{inc}=0$, ~$z_r=0.2$, ~$\beta = 0.8$. Shorter-wavelength emission, such as the $u$ band, originates predominantly from the hotter inner regions of the accretion disk and therefore responds to the central driving variability at shorter time delays. Longer-wavelength emission arises from cooler, larger disk radii, producing broader transfer functions with longer characteristic lags. This wavelength-dependent response encodes the radial temperature structure of the accretion disk.
    }
    \label{fig:transfer function}
\end{figure}

After integrating over solid angle, the time-dependent flux spectrum can be written as
\begin{align}
f_{\nu}(t)
=
\bar{f}_{\nu}
+
\Delta f_{\nu}
\int_0^{\infty}
\Psi_{\nu}(\tau|\lambda)\,
X\bigl(t-\tau\bigr)\,d\tau .
\label{transfer_conv}
\end{align}
Here, $\Psi(\tau|\lambda)$ is the wavelength-dependent transfer function, shown in Fig.~\ref{fig:transfer function}. It describes the delayed response of the disk emission at a given wavelength to variability in the central X-ray source. The transfer function is calculated as
\begin{align}
\Psi_{\nu}\bigl(\tau \mid \lambda\bigr)
= \frac{1}{\Delta f_{\nu}}
  \int_{\Omega}
    \Delta B_{\nu}\bigl(R'\bigr)\,
    \delta\!\bigl(t - \tau\bigr)\,
  d\Omega' .
\end{align}
The mean flux $\bar{f}_{\nu}$ and the variability amplitude $\Delta f_{\nu}$ are given by
\begin{align}
    \bar{f}_{\nu}
    &= \int_{\Omega} \bar{B}_{\nu}(R)\,d\Omega,\\
    \Delta f_{\nu}
    &= \int_{\Omega} \Delta B_{\nu}(R)\,d\Omega .
\end{align}

The transfer function $\Psi_\nu(\tau|\lambda)$, the mean flux $\bar{f}_{\nu}$, and the variability amplitude $\Delta f_{\nu}$ all encode information about the geometry and thermal structure of the accretion disk. Therefore, by modeling the wavelength-dependent variability of AGN light curves, one can infer physical properties of the SMBH accretion disks. 

To generate the simulated spectra, we sample the accretion-disk parameters uniformly within the ranges listed in Table~\ref{tab:simulation params}. For each parameter set, we compute the corresponding transfer function, mean flux $\bar{f}_{\nu}$, and flux-variability amplitude $\Delta f_{\nu}$. We then convolve the DRW driving signal with the transfer function and combine it with $\bar{f}_{\nu}$ and $\Delta f_{\nu}$ to obtain the time-dependent flux spectrum, as described in Eq.~\ref{transfer_conv}.

One of the main goals of our attentive neural process model is to perform Bayesian inference on four parameters that characterize the AGN system and its variability: the black hole mass, $M_{\mathrm{bh}}$; the Eddington ratio, $\lambda_{\mathrm{Edd}}$; the redshift, $z_r$; and the DRW timescale, $\tau$ of the driving variability. In the following section, we describe how the latent representations learned from the simulated light curves are used to infer these parameters.

\begin{table}[tb]
    \centering
    \begin{tabular}{cccccccccc}
        \hline
         & $\log(\tau)$ & $a$ & $\beta$ & $H$ & $\theta_{inc}$ & $z_r$ & $\lambda_{Edd}$ & $\log(M_{bh})$ \\
        \hline
         Min. & -0.5    & -1    & 0.3  & 10.0    & 0.0    & 0.1    & 0.01     & 7.0     \\
         Max & 3.5     & 1    & 1.0   & 50.0    & $80^{\circ}$    & 6.0    & 0.3     & 10.0     \\
        \hline
    \end{tabular}
    \caption{Ranges of the black hole parameters used for simulation. First row is the minimum and the second row is the maximum. We set the prior distribution for all the parameters to be uniform.}
    \label{tab:simulation params}
\end{table}

%\begin{figure}
%    \centering
%    \includegraphics[width=1.0\linewidth]{LSST_filters.png}
%    \caption{The LSST filter responses to \textit{u,g,r,i,z,y} bands.
%    }
%    \label{fig:filter_response}
%\end{figure}
\subsection{LSST Survey Simulation} We convolve the spectra simulated in \S~\ref{sec:Transfer_func} with the LSST filter response curves to obtain light curves in the six observed optical bands, \textit{ugrizy}. To construct realistic mock observations, we further sample the simulated light curves according to the LSST observing cadence from \href{https://survey-strategy.lsst.io/baseline/changes.html}{\texttt{baseline\_v3.4\_10yrs.db}}. For each simulated observation, we assign a photometric uncertainty following the LSST signal-to-noise prescription described in SMTN-002 \citep{SMTN002}. In this formalism, the signal-to-noise ratio is computed from the source counts, sky background, instrumental noise, gain, and the effective number of pixels contributing to the source measurement: \begin{align} \mathrm{SNR} = \frac{C} { \sqrt{ C/g + \left(B/g+\sigma_{\mathrm{instr}}^2\right)n_{\mathrm{eff}} } }, \end{align} where $C$ is the total source counts, $B$ is the sky background counts per pixel, $g$ is the gain, $\sigma_{\mathrm{instr}}$ is the instrumental noise per pixel, and $n_{\mathrm{eff}}$ is the effective number of pixels in the source aperture or PSF footprint. The magnitude uncertainty is then obtained by propagating the flux uncertainty into magnitude space. Since \begin{align} m = -2.5\log_{10}F + \mathrm{constant}, \end{align} and $\mathrm{SNR}=F/\sigma_F$, the corresponding magnitude error is \begin{align} \sigma_m = \frac{2.5}{\ln 10}\frac{1}{\mathrm{SNR}} \approx \frac{1.0857}{\mathrm{SNR}} . \end{align} We then generate the mock observed magnitude by adding Gaussian observational noise: \begin{align} m_{\mathrm{obs}} \sim \mathcal{N} \left( m_{\mathrm{true}},\, \sigma_m^2 \right), \end{align} where $m_{\mathrm{true}}$ is the noiseless simulated magnitude evaluated at the LSST observing epoch. This procedure assigns each observation a noise level determined by the source brightness and the observing conditions of the corresponding LSST visit.

For model training and validation, we apply an additional post-selection criterion based on the signal-to-noise ratio%, which is defined by flux divided by flux error. 
Specifically, we retain only light curves with signal-to-noise ratio $\mathrm{SNR}>2$. This selection removes very low-quality light curves whose variability signal is dominated by observational noise, improving numerical stability and enabling more efficient model training.

Since the model is trained on LSST cadence-sampled light curves, real observational data can be incorporated into the training set in the future without requiring changes to the model architecture. This provides a natural pathway to fine-tune our model on real observational data.

\section{Model Architecture} \label{sec:modeling}
Inspired by the work of \cite{park2021inferringblackholeproperties}, we introduce an ANP network for modeling AGN light curves. In their study, the model was trained and validated only on clean (i.e., noiseless) DRW signals. We extend this approach by applying the model to more realistic simulated LSST light curves, which incorporate wavelength-dependent transfer functions, the LSST observing cadence, and photometric noise. To handle the irregular sampling of LSST light curves and improve numerical stability during training, we tokenize the light-curve data before feeding them into the ANP model, and upgrade their ANP network structure as described in \S~\ref{subsec:data_tokenization} and \S~\ref{subsec: ANP_structure}.

\subsection{Data Tokenization}\label{subsec:data_tokenization}
\begin{figure*}[t]
    \centering
    \includegraphics[width=\textwidth]{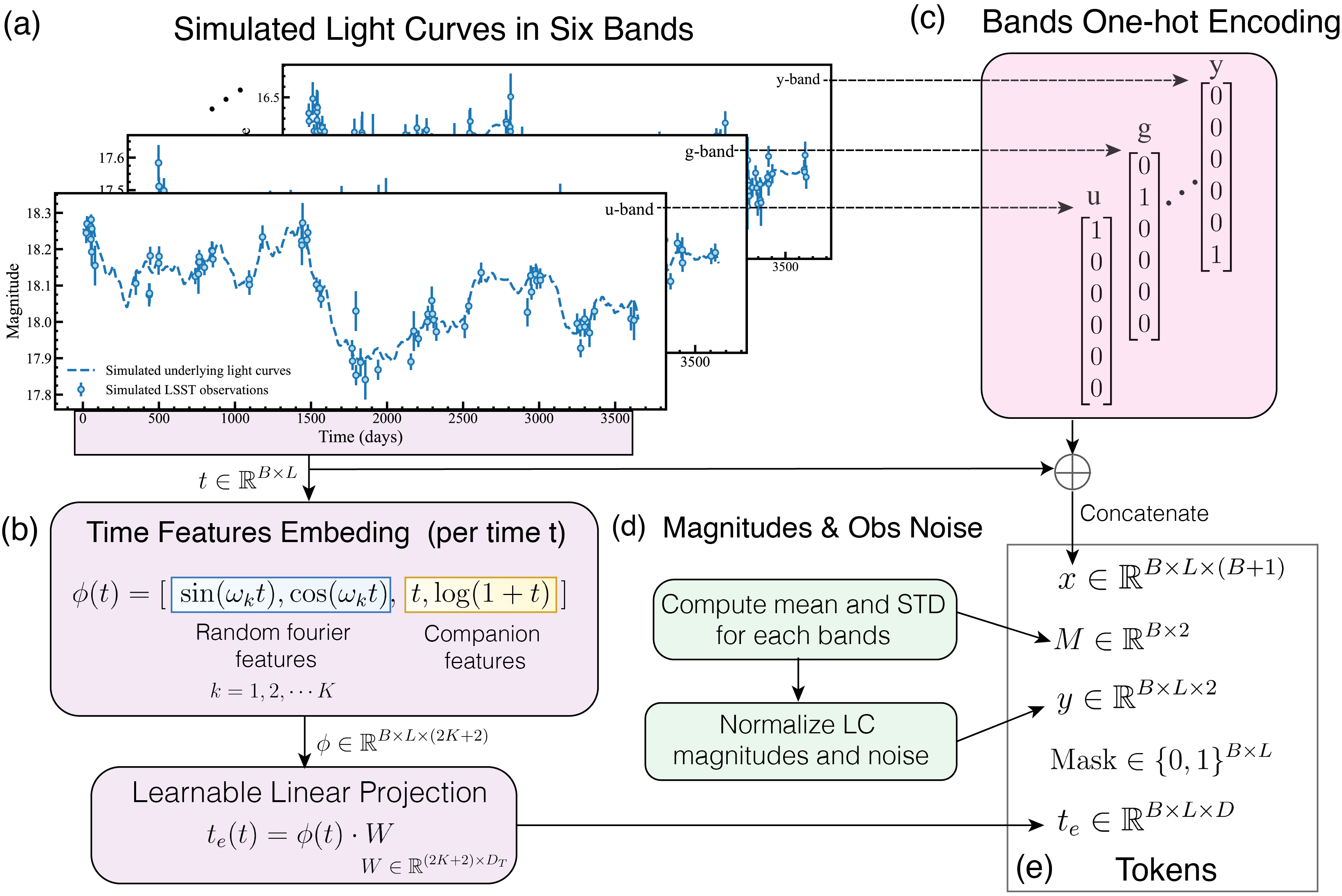}
    \caption{
    Preprocessing pipeline for simulated multi-band LSST light curves.
    \textbf{(a)} Simulated light curves in six LSST bands. Blue points denote simulated LSST observations used as model inputs during training and inference, while blue dashed curves show the simulated underlying light curves, which are used only as ground truth for testing or model evaluation and are not visible to the model.
    \textbf{(b)} Time-feature embedding. Each time step $t$ is mapped to $\phi(t)$ using random Fourier features, $\sin(\omega_k t)$ and $\cos(\omega_k t)$, together with companion features, $t$ and $\log(1+t)$. A learnable linear projection with weight matrix $W$ then maps $\phi(t)$ to the time embedding $t_e$.
    \textbf{(c)} Band identities are encoded as six-dimensional one-hot vectors and concatenated with time $t$ to construct the input tensor $x$.
    \textbf{(d)} For each band, observed magnitudes $m^{(i)}_t$ and observational uncertainties $\epsilon^{(i)}_t$ are normalized using the band-specific magnitude mean $\mu_i^m$ and standard deviation $\sigma_i^m$. The normalized magnitudes and uncertainties form $y$, while the normalization statistics are stored in $M$.
    \textbf{(e)} The preprocessed tensors $x$, $y$, $t_e$ and $M$, together with a binary mask indicating observed entries, are used for model training and inference.
    }
    \label{fig:preprocess}
\end{figure*}

Fig.~\ref{fig:preprocess} summarizes the tokenization pipeline applied to each simulated multi-band light curve before it is passed to the ANP model. Each light curve is tokenized into:
$$\mathcal{D}=\{t_e, x, y, M, \mbox{Mask}\}$$, where $t_e\in\mathbb{R}^{B\times L}$, $x\in\mathbb{R}^{B\times L\times(B+1)}$, $y\in \mathbb{R}^{B\times L\times 2}$,  $M\in \mathbb{R}^{B\times L\times 2}$ and $\mbox{Mask}\in\{0,1\}^{B\times L}$.

We first construct a time embedding for the observation times. For each light curve, the time tensor is denoted by $t\in\mathbb{R}^{B\times L}$, where $B=6$ is the number of LSST photometric bands and $L$ is the maximum sequence length. Each time value is mapped to a feature vector
\begin{align}
    \phi(t)
    =
    \left[
    \sin(\omega_k t),\,
    \cos(\omega_k t),\,
    t,\,
    \log(1+t)
    \right],
    \quad k=1,\dots,K 
\end{align}
so that $\phi(t)\in\mathbb{R}^{B\times L\times(2K+2)}$. This embedding combines sinusoidal Fourier features with linear and logarithmic time features.

The sinusoidal Fourier features are constructed by concatenating $\sin(\omega_k t)$ and $\cos(\omega_k t)$ over a set of frequencies $\{\omega_k\}_{k=1}^{K}$. We tuned the frequency range to cover the relevant temporal scales of the simulated LSST survey. The lowest frequency corresponds to the full observing baseline, $\omega_1=2\pi/3000~{\rm day}^{-1}$, while the highest frequency corresponds to the shortest cadence-resolvable timescale, $\omega_K=2\pi/1~{\rm day}^{-1}$. These Fourier features allow the model to represent temporal variations across multiple characteristic timescales. This is particularly important for irregularly sampled LSST light curves, because the model receives a richer representation of relative phase and temporal separation rather than relying only on the raw observation times.

The linear time feature, $t$, preserves the absolute temporal ordering of the observations across the survey baseline. This helps remove the ambiguity introduced by purely periodic features, which can assign similar phases to widely separated times. The logarithmic feature, $\log(1+t)$, compresses large time values while retaining sensitivity to early-time variations. This improves numerical conditioning and helps the model represent variability over the broad range of timescales expected in AGN light curves, from days to years.

After constructing $\phi(t)$, we apply a learnable linear projection to map the time features into the time-embedding space to get token $t_e$, which will later be used by the ANP model:
\begin{align}
    t_e(t) = \phi(t) W ,
\end{align}
where $W\in\mathbb{R}^{(2K+2)\times D_T}$ is a trainable weight matrix and $D_T$ is the dimension of the final time embedding. This projection allows the model to learn the most informative combinations of Fourier, linear, and logarithmic time features directly from the training data. The resulting time embedding $t_e\in\mathbb{R}^{B\times L\times D_T}$ is then used as the temporal representation of the light curve in the downstream ANP model.

Since the LSST observing cadence samples different photometric bands irregularly, the representation used by \cite{park2021inferringblackholeproperties}, in which the magnitudes from all six bands are combined into a single six-dimensional vector at each time step, is not directly applicable here. For sparsely and asynchronously sampled LSST light curves, this representation would introduce many empty entries. If these missing observations were filled with zeros, the model could incorrectly interpret them as true magnitude measurements rather than unobserved data.

Instead, we represent each observation as an individual token and encode the corresponding photometric band using a one-hot vector. The six LSST bands, $u,g,r,i,z,y$, are each mapped to a six-dimensional one-hot band vector, as illustrated in Fig.~\ref{fig:preprocess}(c). This band-identity vector is concatenated with the observation time $t$ to form the input token   $ x \in \mathbb{R}^{B\times L \times (B+1)} $,
where, again, $B=6$ is the number of LSST bands and $L$ is the maximum sequence length after padding.

For each observation in band $j$ at time $t$, we denote the observed magnitude and observational uncertainty as $m_t^{(j)}$ and $\epsilon_t^{(j)}$, respectively. To improve numerical stability during training, we normalize the magnitudes and uncertainties in each band independently for all training and testing data:
\begin{align}
    \bar{m}_t^{(j)}
    &=
    \frac{m_t^{(j)}-\mu_j^m}{\sigma_j^m},\\
    \bar{\epsilon}_t^{(j)}
    &=
    \frac{\epsilon_t^{(j)}}{\sigma_j^m}.
\end{align}
Here, $\mu_j^m$ and $\sigma_j^m$ are the mean and standard deviation of the observed magnitudes in band $j$ of a sample AGN:
\begin{align}
    \mu_j^m
    &=
    \frac{1}{N_j}\sum_{t=1}^{N_j} m_t^{(j)},\\
    \sigma_j^m
    &=
    \sqrt{
    \frac{1}{N_j-1}
    \sum_{t=1}^{N_j}
    \left(m_t^{(j)}-\mu_j^m\right)^2
    },
\end{align}
where $N_j$ is the number of observations in band $j$.

The normalized magnitude and uncertainty are concatenated to form the output token
\begin{align}
    y = [\bar{m}_t^{(j)},\bar{\epsilon}_t^{(j)}] \in \mathbb{R}^{B\times L \times 2}.
\end{align}
Since this normalization removes information about the absolute brightness and variability amplitude of each band, we also store the normalization statistics,
\begin{align}
    M_j = [\mu_j^m, \sigma_j^m],
\end{align}
as an additional input to the model. This allows the model to retain information about both the absolute magnitude scale and the band-dependent variability strength for each light curves.

Finally, to enable batch training, all light curves are padded along the sequence dimension to the same length $L$. We use binary masks to distinguish real observations from padded entries, ensuring that the model does not treat padded zeros as physical measurements.

\subsection{Attentive Neural Processing 
Model}\label{subsec: ANP_structure}
\begin{figure*}[t]
    \centering
    \includegraphics[width=\textwidth]{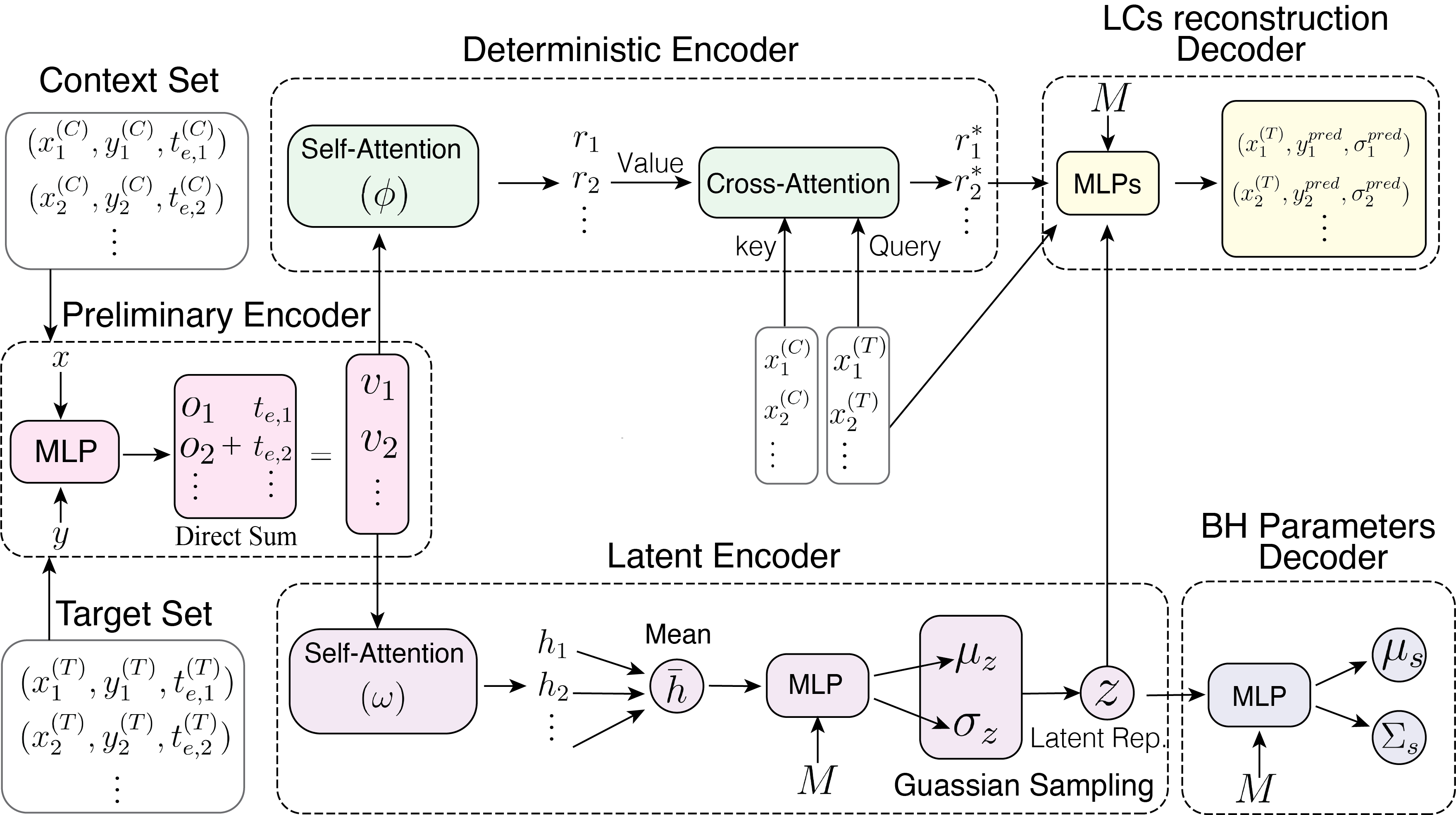}
    \caption{The architecture of the attentive neural process. The preliminary encoder maps each observation $(x_j, y_j, \sigma_{y,j})$ through an MLP to an embedding $o_j$, which is summed with a learned time embedding $t_{e,j}$ to form the point representation $v_j$. In the latent path, masked multi-head self-attention over ${v_j}$ produces ${h_j}$, which are mean-pooled into a set-level summary $\bar{h}$. Together with $M$ — the per-band mean and variance statistics of the observed light curve — $\bar{h}$ is passed through an MLP that outputs the parameters $(\mu_z, \sigma_z)$ of a Gaussian over the latent variable, from which $z$ is sampled. This latent representation serves the downstream heads: black-hole parameter inference and light-curve reconstruction. In the deterministic path, a second masked multi-head self-attention yields point representations $r_j$, which act as values in a masked multi-head cross-attention with the observed times $x^{(C)}$ as keys and the prediction times $x^{(T)}$ as queries; the attention weights thus form a learned, data-adaptive interpolation kernel over the light curve's temporal correlation structure, producing a target-specific representation $r^*_k$ for each prediction time. The light-curve reconstruction decoder consists of a shared trunk and two output branches: the trunk MLP combines $(r^*_k, z, x^{(T)}_k, M)$ and yields the mean prediction $\mu^{\mathrm{pred}}_k$, while the standard-deviation branch additionally receives distance features $\Delta t_k$ — the temporal separation between $x^{(T)}_k$ and the nearest observation — and outputs $\sigma^{\mathrm{pred}}_k$, allowing the predictive uncertainty to widen inside observing gaps and contract near data. The black-hole parameter decoder maps $(z, \bar{h}, M)$ to the mean vector and Cholesky factor of a full-rank four-dimensional Gaussian posterior over $(\log M_{\mathrm{BH}}, \log\lambda_{\mathrm{edd}}, z_r, \log\tau)$. All attention operations are masked so that padded entries never contribute.}
    \label{fig:ANP_structure}
\end{figure*}

Here we present the ANP model, following the frameworks of \cite{park2021inferringblackholeproperties} and \cite{dubois2020npf}. %The goal of the model is to learn a compact latent representation of LSST-like AGN light curves.

As shown in Fig.~\ref{fig:ANP_structure}, the model consists of five main components: a preliminary encoder, a deterministic encoder, a latent encoder, a decoder for light-curve reconstruction, and a decoder for black hole parameter inference. The modular structure of the model also allows additional downstream tasks, such as anomaly detection or binary AGN classification, to be incorporated in future work.

During training, each simulated LSST light curve is divided into a context set and a target set, denoted by
\begin{align}
    \mathcal{C} &= \left(x^{(C)}, y^{(C)}, t_e^{(C)}\right),\\
    \mathcal{T} &= \left(x^{(T)}, y^{(T)}, t_e^{(T)}\right),
\end{align}
where $x$ contains the time and band information, $y$ contains the normalized magnitude and observational uncertainty, and $t_e$ is the learned time embedding described in \S~\ref{subsec:data_tokenization}. Following \cite{park2021inferringblackholeproperties}, we choose the context set to be a subset of the target set, $\mathcal{C} \subset \mathcal{T}$. In this setup, the context set provides the observed conditioning points used by the model, while the target set defines the points at which the model is trained to reconstruct the light curve and evaluate the predictive distribution. Details on the context and target set selection can be found in \S~\ref{sec:training_detail}.

Both the context and target observations are first passed through a preliminary encoder. This encoder maps the tokens $x$ and $y$ into an initial feature representation using a multilayer perceptron (MLP). The resulting feature vector is then combined with the learned time embedding $t_e$ through element-wise addition:
\begin{align}
    v = \mathrm{MLP}(x,y) + t_e .
\end{align}
This operation produces a sequence of preliminary representations $v$, which encode the photometric information, observational uncertainty, band identity, and temporal position of each data point. In Fig.~\ref{fig:ANP_structure}, the direct addition of the time embedding is analogous to the use of positional encoding in sequence models. It allows temporal information to be incorporated into the representation at an early stage, while keeping the dimensionality fixed for the subsequent attention layers. The resulting representations are then used as inputs to the deterministic and latent encoders, which learn the temporal correlation information of the light curve conditioned on the available context observation and form the latent representations $z$.

In the latent encoder block, we use the sequence of preliminary representations $v$ to infer a global latent variable that summarizes the overall structure of the light curve. We first apply a masked multi-head self-attention layer to $v$:
\begin{align}
    h = \mathrm{SelfAttn}_{\omega}(v),
\end{align}
where $\omega$ denotes the parameters of the latent masked multi-head self-attention module. The masked multi-head self-attention operation allows each observation to attend to all other observations in the light curve, thereby capturing temporal dependencies and correlations across different time and photometric bands. In this sense, the learned attention weights provide a flexible, data-driven analogue of an autocorrelation structure, but with the advantage of capturing longer correlations and cross-band photometric correlation information.

The output representations $h$ are then aggregated by taking their mean over the context/target sequence dimension:
\begin{align}
    \bar{h}^ = \frac{1}{L}\sum_{j=1}^{L} h_j ,
\end{align}
where $L$ is the length of context/target set. This aggregation produces a permutation-invariant summary of the context light curve. We then concatenate $\bar{h}$ with the normalization statistics $M$, which contain the band-dependent magnitude means and standard deviations. Including $M$ allows the latent encoder to retain information about the absolute brightness scale and variability amplitude that is removed during normalization.

The concatenated representation is passed through an MLP to parameterize a Gaussian latent distribution for both context and target set:
\begin{align}
    q^{(C)}(z^{(C)}|\mathcal{C}) =
    \mathcal{N}\left(\mu^{(C)}_z, \sigma_z^{2(C)}\right),
\end{align}
\begin{align}
    q^{(T)}(z^{(T)}|\mathcal{T}) =
    \mathcal{N}\left(\mu^{(T)}_z, \sigma_z^{2(T)}\right)
\end{align}
where
\begin{align}
    \mu^{(C)}_z, \sigma^{(C)}_z = \mathrm{MLP}(\bar{h}^{(C)}, M)
    \label{latent_context}
\end{align}
\begin{align}
     \mu^{(T)}_z, \sigma^{(T)}_z = \mathrm{MLP}(\bar{h}^{(T)}, M)
    \label{latent_target}
\end{align}
are output from the MLP given the context or target set.
Latent representation $z^{(C)}$ and $z^{(T)}$ are then sampled from these distributions using the reparameterization trick. This latent variable provides a compact probabilistic summary of the input light curve and is shared by the downstream decoders for light-curve reconstruction and black hole parameter inference. 

The latent representation inferred from the context set, $z^{(C)}$, is used for black hole parameter inference and light-curve reconstruction. For black hole parameter inference, we concatenate $z^{(C)}$ with the normalization statistics $M$. The combined representation is then passed to a black hole parameter decoder, implemented as an MLP:
\begin{align}
    \mu_s, \Sigma_s = \mathrm{MLP}_{\mathrm{BH}}\left(z^{(C)}, M\right).
\end{align}
The decoder outputs the mean vector $\mu_s$ and covariance matrix $\Sigma_s$ of a multivariate Gaussian distribution over the target physical parameters,
\begin{align}
    p(s|z^{(C)},M) = \mathcal{N}\left(\mu_s, \Sigma_s\right),
\end{align}
where $s$ denotes the set of black hole and variability parameters to be inferred, including $M_{\mathrm{bh}}$, $\lambda_{\mathrm{Edd}}$, $z_r$, and $\tau$ in this research. This probabilistic output allows the model to estimate both the most likely parameter values and their associated uncertainties.

In the deterministic encoder, the model learns a target-specific representation of the light curve conditioned on the context observations. We first apply a masked multi-head self-attention layer to the preliminary context representations $v^{(C)}$:
\begin{align}
    r = \mathrm{SelfAttn}_{\phi}\left(v^{(C)}\right),
\end{align}
where $\phi$ denotes the parameters of the deterministic self-attention module. This masked multi-head self-attention operation allows the model to capture temporal correlations and cross-band dependencies among the observed context points.

We then apply a multi-head cross-attention layer to transfer information from the context set to the target locations. In this layer, the self-attention output from context set $r$ serves as the value, the context inputs $x^{(C)}$ serve as the keys, and the target inputs $x^{(T)}$ serve as the queries:
\begin{align}
    r^* = \mathrm{CrossAttn}\left(
    Q=x^{(T)},\,
    K=x^{(C)},\,
    V=r
    \right).
\end{align}
The resulting representation $r^*$ provides a deterministic summary of the context light curve at each target location (i.e., for a specific time and filter). Because the queries are defined at the target points, the cross-attention layer allows the model to infer which context observations are most relevant for predicting each target observation. 

The deterministic representation $r^*$, together with the latent variable
$z^{(C)}$ inferred from Eq.~\ref{latent_context}, is passed to the light-curve
reconstruction decoder, which predicts the light-curve magnitude distribution
at the target times and filters. The decoder consists of a shared trunk and
two output branches. The trunk MLP combines the target-specific deterministic
representation with the latent variable, the target input location, its time
embedding, and the per-band summary statistics $M$,
\begin{align}
    u_j = \mathrm{MLP}_{\mathrm{trunk}}
    \left(r^*_j,\, z^{(C)},\, x^{(T)}_j,\, t_{e,j},\, M\right),
\end{align}
from which the mean prediction is obtained through a linear head,
$\mu_{\mathrm{pred},j} = W_\mu u_j$. The standard deviation is produced by a
dedicated branch that additionally receives distance features $\Delta t_j$
--- the temporal separation between $x^{(T)}_j$ and the nearest observation, in linear and logarithmic form:
\begin{align}
    \sigma_{\mathrm{pred},j} = \mathrm{MLP}_{\sigma}
    \left(u_j \oplus \Delta t_j\right),
    \label{dist_sig}
\end{align}
with positivity enforced by a bounded transform. Decoupling the uncertainty
branch from the mean prediction, and conditioning it explicitly on the
distance to the data, allows the predictive width to contract near
observations and grow inside seasonal gaps without perturbing the mean
reconstruction.

The decoder thus defines a Gaussian predictive distribution for the
underlying light curve at each target location,
\begin{align}
    p\left(y^{(T)}_j \mid x^{(T)}_j, r^*_j, z^{(C)}\right)
    =
    \mathcal{N}\left(\mu_{\mathrm{pred},j},\,
    \sigma_{\mathrm{pred},j}^2\right).
\end{align}
During training, the likelihood is evaluated only against observed
photometry, whose reported measurement uncertainty $\bar{\epsilon}_j^2$ is added in
quadrature,
\begin{align}
    p\left(y^{(T)}_{\mathrm{obs},j} \mid \cdot\right)
    =
    \mathcal{N}\left(\mu_{\mathrm{pred},j},\,
    \sigma_{\mathrm{pred},j}^2 +  \bar{\epsilon}_j^2\right),
\end{align}
so that $\sigma_{\mathrm{pred}}$ learns to represent the model's intrinsic
predictive uncertainty about the underlying light curve rather than the
photometric noise. This probabilistic formulation allows the model to
reconstruct the light curve while providing calibrated
uncertainties at every target location.

\subsection{Loss Function}

Following \cite{park2021inferringblackholeproperties}, we train the model by
maximizing the evidence lower bound (ELBO), or equivalently by minimizing the
corresponding negative ELBO loss, augmented with a regularization term that
anchors the model's behavior inside observing gaps. The total loss contains
four terms: the negative log likelihood of the light curve
$\mathcal{L}_{\mathrm{LC}}$, the negative log likelihood of the black hole
parameters $\mathcal{L}_{\mathrm{BH}}$, the KL divergence between the latent
distributions inferred from the target and context sets
$\mathcal{L}_{\mathrm{KL}}$, and the gap-anchor regularizer
$\mathcal{L}_{\mathrm{gap}}$:
\begin{align}\label{eq:loss}
    \mathcal{L}
    =
    W_y \mathcal{L}_{\mathrm{LC}}
    +
    \mathcal{L}_{\mathrm{BH}}
    +
    \beta \,\mathcal{L}_{\mathrm{KL}}
    +
    W_{\mathrm{gap}}\, \mathcal{L}_{\mathrm{gap}} .
\end{align}
The hyperparameter $W_y$ controls the relative weight of the light-curve
reconstruction term. Increasing $W_y$ places greater emphasis on accurately reconstructing the target light curves, while decreasing it gives relatively more weight to black hole parameter inference and latent-space regularization. The weight $\beta$ controls the strength of the latent regularization relative to the reconstruction terms, in the spirit of the $\beta$-VAE framework \citep{higgins2017betavae}.

For the light-curve reconstruction term, the decoder predicts a Gaussian distribution at each target point with mean $y_{\mathrm{pred},j}$ and model variance $\sigma_{\mathrm{pred},j}^2$. Since the target observations also contain LSST photometric noise, we add the observational variance $\bar{\epsilon}_j^2$ in quadrature. Thus, for each target point $j$,
\begin{align}
    y_j^{(T)}
    \sim
\mathcal{N}\left(\mu_{\mathrm{pred},j},\,
    \sigma_{\mathrm{pred},j}^2 +  \bar{\epsilon}_j^2\right)
\end{align}
The corresponding negative log likelihood is
\begin{align}
    \mathcal{L}_{\mathrm{LC}}
    =
    \frac{1}{2}
    \sum_{j\in\mathcal{T}}
    \left[
    \log\left(2\pi s_j^2\right)
    +
    \frac{
    \left(y_j^{(T)}-y_{\mathrm{pred},j}\right)^2
    }{
    s_j^2
    }
    \right]
    \qquad
\end{align}
\begin{align}
    s_j^2 =
\sigma_{\mathrm{pred},j}^2 + \bar{\epsilon}_j^2
\end{align}
where the sum is taken over the observed target points after masking padded entries. Here, $\bar{\epsilon}_j$ denotes the normalized LSST observational uncertainty at the $j$-th target point.

For black hole parameter inference, the parameter decoder predicts a multivariate Gaussian distribution with mean $\mu_s$ and covariance matrix $\Sigma_s$:
\begin{align}
    \mathbf{s}
    \sim
    \mathcal{N}
    \left(
    \mu_s,
    \Sigma_s
    \right),
\end{align}
where $\mathbf{s}$ is the true parameter vector, containing the black hole mass, Eddington ratio, redshift, and damped-random-walk timescale. The negative log likelihood is therefore
\begin{align}
    \mathcal{L}_{\mathrm{BH}}
    =
    \frac{1}{2}
    \left[
    \left(\mathbf{s}-\mu_s\right)^{\mathsf{T}}
    \Sigma_s^{-1}
    \left(\mathbf{s}-\mu_s\right)
    +
    \log|\Sigma_s|
    +
    d_s \log(2\pi)
    \right],
\end{align}
where $d_s=4$ is the dimension of the inferred parameter vector.

The KL term regularizes the latent space. Writing $q^{(C)}$ and $q^{(T)}$ for
the latent distributions inferred from the context and the full target set,
\begin{align}
    \mathcal{L}_{\mathrm{KL}}
    =
    \sum_{d=1}^{D}
    \max\!\left[
    \lambda_{\mathrm{fb}},\,
    \mathrm{KL}\!\left(
    q^{(T)}_d\left(z_d \mid \mathcal{T}\right)
    \,\Vert\,
    q^{(C)}_d\left(z_d \mid \mathcal{C}\right)
    \right)
    \right],
\end{align}
where the KL divergence is accumulated per latent dimension $d$. Here we clamp the KL divergence of each dimension to some value $\lambda_{\mathrm{fb}}$. The clamping ensures the minimum information capacity per dimension and prevents the posterior collapse to the prior
\citep{kingma2017improvingvariationalinferenceinverse}. The weight $\beta$ for the KL divergence is increased linearly from
$0$ to $1$ over the first $20$ training epochs; this is a standard technical ("annealing") that allows the latent variable to encode information about the light curve before the regularization toward
the context distribution takes full effect. Throughout training the decoders
only condition on context latent $z^{(C)}$, so the target-set encoding enters the
objective only through this KL term. At deployment, where the full observed
light curve constitutes the context, the two distributions coincide by
construction. Here we adopt $\lambda_{\mathrm{fb}} = 0.08$ per dimension.

The final term, $\mathcal{L}_{\mathrm{gap}}$, addresses a failure mode specific to seasonally gapped
cadences. At epochs far from any observation the likelihood provides no
meaningful constraint. We find that, without explicit regularization, the predictive mean can
develop spurious structure with unreliable uncertainties. We
therefore introduce \emph{anchor times} or query times drawn uniformly at
random over the observing baseline. At said anchor times, the decoder is asked to
predict flux measurements even where no observed measurements exist. These predictions are never used in the reconstruction likelihood. %but which carry no photometric measurement --- they are pure
%evaluation locations, contributing nothing to the likelihood, and the %true
%underlying light curve is never consulted. 
Instead, at each training step we draw
$R = 64$ random anchor times per light curve, evaluate the decoder there in every
band to obtain $(\mu_{\mathrm{pred},k}, \sigma_{\mathrm{pred},k})$, and
penalize departures from the AGN's mean flux and an uncertainty prediction that significantly deviates from the intrinsic variability amplitude  $\sigma_\infty$:
%only statistically defensible prediction for
%a mean-reverting process with no nearby data: a mean reverting to the band
%average (zero, in normalized units) and a width approaching the band's
%intrinsic variability amplitude $\sigma_\infty$,
\begin{align}
    \mathcal{L}_{\mathrm{gap}}
    =
    \frac{1}{R}\sum_{k=1}^{R}
    \Big[
    &w_\mu(\Delta t_k)\, \mu_{\mathrm{pred},k}^2
    \nonumber\\
    &+
    w_\sigma(\Delta t_k)
    \left(\log \sigma_{\mathrm{pred},k} - \log \sigma_\infty\right)^2
    \Big],
\end{align}
where $\Delta t_k$ is the separation between anchor $k$ and the nearest observation, and
$\sigma_\infty^2 = \max\!\left(1 - \langle\bar{\epsilon}^2\rangle,\,
\sigma_{\min}^2\right)$ estimates the intrinsic variability variance of
the normalized band from its observed scatter and mean photometric
variance. The gates $w_\mu$ and $w_\sigma$ confine the penalty to
genuinely unconstrained epochs, so that anchors falling near observations
are unaffected by construction. The width gate
$w_\sigma(\Delta t) = 1 - e^{-\Delta t / \tau_g}$ with $\tau_g = 45$~days
activates smoothly away from data, while the mean term uses a more
conservative gate that is identically zero within $60$~days of an
observation and ramps to unity over the following $90$~days, ensuring
that intra-season predictions are never pulled toward the band mean.
$\mathcal{L}_{\mathrm{gap}}$ is thus a prior-consistency penalty only,
independent of any data at the anchor locations. We adopt
$W_{\mathrm{gap}} = 0.15$. The gap regularization term basically assumed that the process itself is stationary and thus it's power spectrum is asymptotically constant. So it's variability behaves the same in the seasonal gaps. 

\subsection{Training Details}\label{sec:training_detail}
Our final model uses a hidden and latent dimension of 64, with two layers in
each of the latent encoder, the deterministic encoder, and the decoder, for a
total of 367{,}953 trainable parameters. All attention modules --- the
self-attention in both encoders and the cross-attention in the deterministic
encoder --- use masked multi-head scaled dot-product attention with four
heads; masking guarantees that padded entries never contribute to any
attention weight. Observation times are encoded with a fixed random Fourier feature embedding
of $K = 64$ frequencies, log-spaced in period from ${\sim}1$~day to the full
observational baseline, supplemented by linear and logarithmic time terms and
linearly projected into the 64-dimensional hidden space.

All MLP blocks use Monte Carlo (MC) dropout, which remains active at
inference time and, combined with sampling of the latent variable,
provides the epistemic contribution to the predictive uncertainty. We use a
channel-dropout rate of 0.05 in the encoder and decoder trunks and 0.15 in
the black hole parameter decoder. Dropout on the attention weights themselves
was evaluated in an ablation and found unnecessary; the final model uses
none.

We set $W_y = 2$ in the loss function (Eq.~\ref{eq:loss}), chosen together
with the latent dimension and number of layers via a grid search on the
validation loss. We train with the Adam optimizer at an initial learning rate
of $10^{-3}$ and a batch size of 20 for 200 epochs (${\approx}30$~h on a
single NVIDIA A100), clipping gradients to a global norm of 2. The black hole
parameter decoder is a distinct optimizer parameter group with a reduced
learning rate of $6\times10^{-4}$ and weight decay of $3\times10^{-4}$, which
we find necessary to balance its convergence against the shared encoders.
The learning rate is halved whenever the validation reconstruction loss fails
to improve for ten consecutive epochs.

During training, each light curve is turned into a prediction episode by
sampling from its observed points. In other words, a random fraction, drawn uniformly between
30\% and 95\%, forms the context set, and with probability 0.7 up to three
contiguous windows, each spanning 5--25\% of the baseline, are additionally
removed from the context to imitate extended data gaps. The full set of
observed points serves as the target. The training, validation, and test
sets contain 50{,}018, 20{,}029, and 8{,}000 simulated light curves,
respectively, generated with disjoint random seeds.

\section{Results} \label{sec:results}
\subsection{Black Hole Parameter Predictions}
\begin{figure}[tb]
    \centering
    \includegraphics[width=1.0\linewidth]{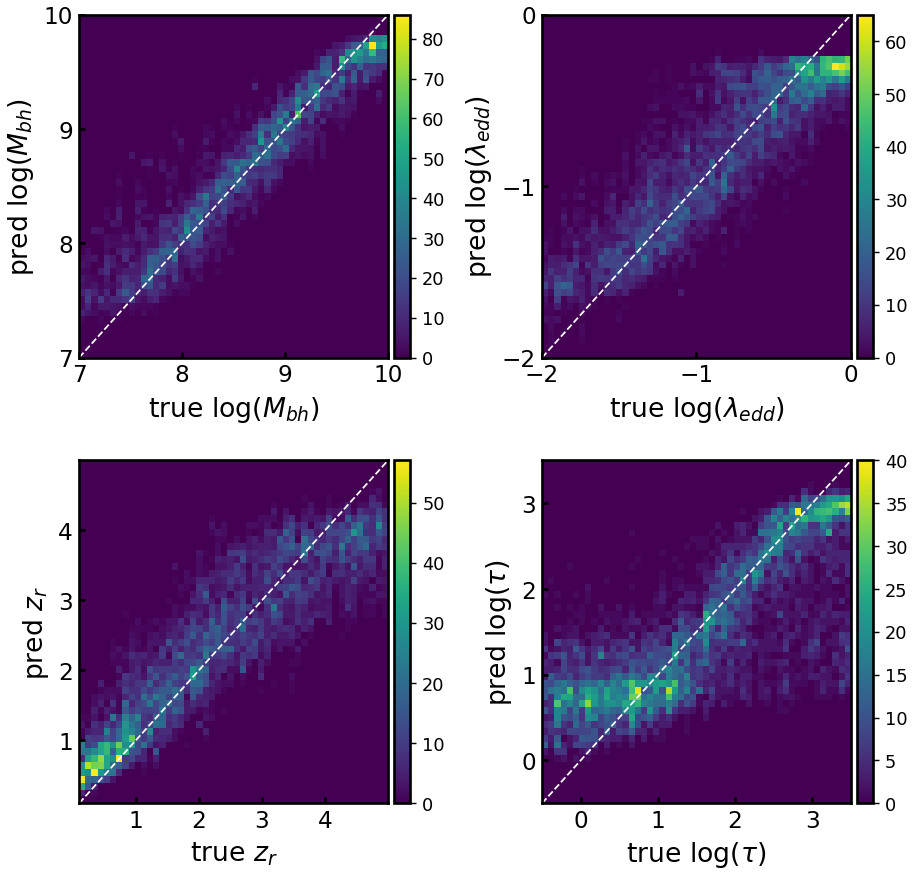}
    \caption{
Model predictions compared with ground-truth values for $\log(M_{\mathrm{bh}})$, $\log(\lambda_{\mathrm{Edd}})$, $z_r$, and $\log(\tau)$ on the testing data set. Each panel shows a two-dimensional histogram, with color indicating the number of objects in each bin. The dashed white line marks the one-to-one relation. The alignment of the distributions with the diagonal indicates that the model captures the dependence of the light curves on the AGN physical and variability parameters. The deviation in $\log(\tau)$ at low $\log(\tau)$, corresponding to short-timescale or high-frequency variability, is likely caused by the finite LSST cadence. Since LSST samples the light curves on timescales of order a few days, variability on shorter or comparable timescales is difficult to resolve and may be affected by aliasing.}
    \label{fig:BH_parameter_pred}
\end{figure}

To demonstrate the effectiveness of the learned latent representation, we infer AGN physical and variability parameters from the latent representation using an MLP decoder. Fig.~\ref{fig:BH_parameter_pred} compares the predicted posterior mean, $\mu_s$, with the ground-truth parameter values for $\approx 8,000$ light curves in the testing set. The white dashed diagonal line indicates the one-to-one relation corresponding to perfect recovery. The predictions for $\log(M_{\mathrm{bh}})$, $\log(\lambda_{\mathrm{Edd}})$, and $z_r$ follow the diagonal, indicating that the latent representation preserves information relevant to the underlying AGN parameters.

The prediction of $\log(\tau)$ shows larger deviations at both small and large values of $\log(\tau)$. At $\log(\tau)<1$ days, corresponding to short-timescale or high-frequency variability, the finite LSST cadence limits the ability to resolve rapid variations and can introduce aliasing effects. At large $\log(\tau)>3$ days, the variability timescale becomes comparable to, or longer than, the survey baseline, making it difficult to constrain the full characteristic timescale from the observed light curve.

\begin{figure}[tb]
    \centering
    \includegraphics[width=1.0\linewidth]{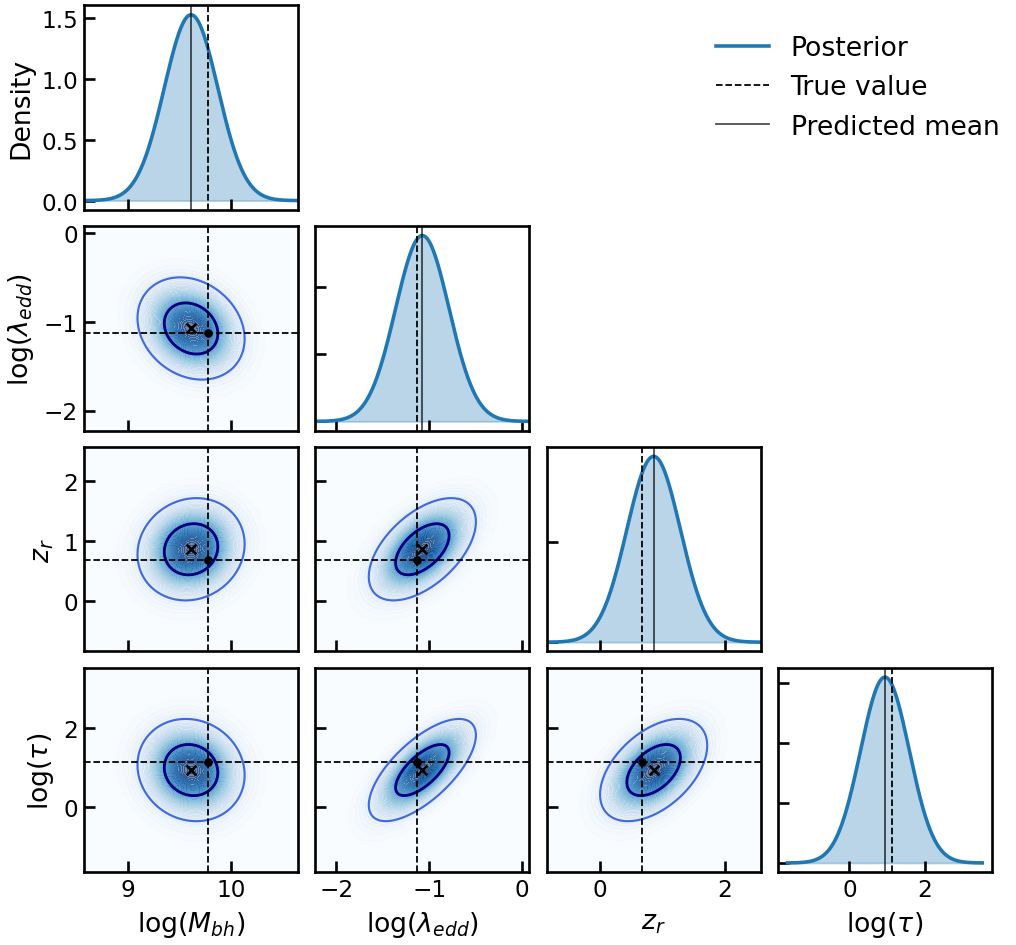}
    \caption{
Example posterior distribution inferred by the ANP model for a single testing light curve. The diagonal panels show the one-dimensional marginal posterior distributions for $\log(M_{\mathrm{bh}})$, $\log(\lambda_{\mathrm{Edd}})$, $z_r$, and $\log(\tau)$, while the off-diagonal panels show the corresponding two-dimensional joint posterior distributions. Dashed black lines denote the ground-truth parameter values, solid gray lines denote the predicted posterior means, the black cross on at the center of the distribution indicates the predicted posterior mean and blue contours indicate the posterior probability density. This corner plot evaluates the performance of the Bayesian inference by comparing the inferred posterior distributions with the ground-truth parameter values.
}
    \label{fig:Coner_plot}
\end{figure}

\begin{figure}[t!]
    \centering
    \includegraphics[width=1.0\linewidth]{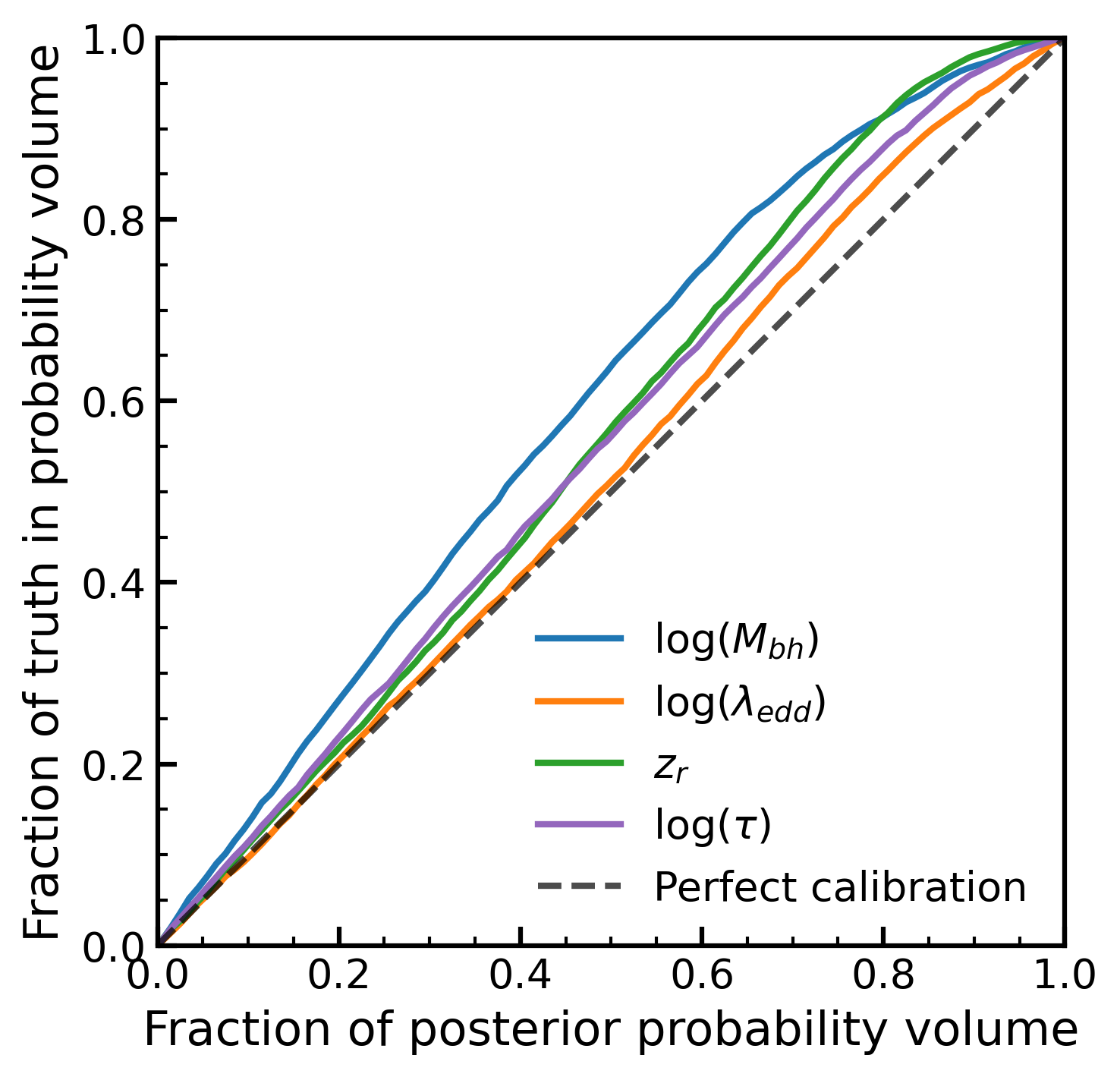}
    \caption{
Calibration curve for the posterior distributions of the inferred AGN host SMBH parameters on the testing set. The horizontal axis shows the fraction of posterior probability volume, and the vertical axis shows the fraction of true parameter values contained within that volume. The dashed black line indicates perfect calibration, for which the credible regions contain the true values at the expected frequency.
} 
    \label{fig:meta_calibration}
\end{figure}

Fig.~\ref{fig:Coner_plot} shows the posterior distribution of the inferred parameters for a representative testing light curve. The diagonal panels show the one-dimensional marginal posterior distributions, while the lower off-diagonal panels show the two-dimensional joint posterior distributions between parameter pairs. All physical parameters are well-constrained with our model. %In this example, the true parameter values lie within the high-probability posterior regions, demonstrating that the model provides not only point estimates but also meaningful uncertainty estimates for the inferred black hole and variability parameters.

We further evaluate the uncertainty calibration of the inferred SMBH
parameters using the coverage calibration curves shown in
Fig.~\ref{fig:meta_calibration}. For each parameter and each nominal
credible level $p$, we compute the fraction of test objects whose
ground-truth value falls within the central $p$ credible interval of the
marginal posterior. Aperfectly calibrated posterior follows the one-to-one
relation shown by the dashed line, curves above it indicate conservative
(over-wide) uncertainties, and curves below it indicate over-confidence.
 At the 68\% (95\%) nominal level the empirical coverages
are $82.5\%$ ($98.7\%$) for $\log(M_{\rm bh})$, $72.2\%$ ($96.9\%$) for
$\log(\lambda_{\rm edd})$, $77.6\%$ ($99.5\%$) for $z_r$, and $75.1\%$
($98.5\%$) for $\log(\tau)$: the credible intervals are modestly wider than
necessary, and no credible level for any parameter is over-confident.

We quantify accuracy with the normalized RMSE of the posterior mean,
defined as the RMSE divided by the width of the corresponding uniform
prior. We obtain $8.2\%$ for $\log(M_{\rm bh})$, $12.0\%$ for
$\log(\lambda_{\rm edd})$, $11.8\%$ for $z_r$, and $16.8\%$ for
$\log(\tau)$. For reference, an estimator that always returns the prior
mean yields $1/\sqrt{12} \approx 28.9\%$ for a uniform prior, so all four
parameters are recovered with substantially more information than the
prior provides.

%The damping timescale admits a direct comparison with the DRW GP baseline,
%which is the only parameter accessible to it
We next investigate the DRW time scale, which is the only parameter that can be directly inferred by both our model and the GP baseline. To obtain the DRW time scale from GP model, we take the
median of the six per-band maximum a posteriori kernel timescales
$\hat{\tau}_{\rm obs}$ and convert to the rest frame using the \emph{true}
redshift. We note that this comparison gives an advantage to the GP model, since our network must
actually infer the redshift itself. On a common subset of test objects, the GP
recovers $\log(\tau)$ with a normalized RMSE of $20.2\%$ and a systematic
bias of $-0.26$~dex --- the characteristic underestimate of
maximum-likelihood timescale fits on baselines comparable to $\tau$ ---
whereas our ANP model achieves $16.0\%$ with a bias of $-0.15$~dex and is
closer to the truth for $61\%$ of individual objects. We attribute the success of our model over the baseline to
the amortized population prior. In other words, because our is trained across $5\times10^4$ light curves,
the network implicitly performs the hierarchical shrinkage that individual object
fits cannot. 
%More fundamentally, the comparison is available only for
%$\tau$: the GP baseline provides no access to $\log(M_{\rm bh})$,
%$\log(\lambda_{\rm edd})$, or $z_r$, and its timescale estimate is a point
%value without an associated posterior, so it offers no notion of
%probability coverage. The joint, calibrated four-parameter posterior is a
%capability of the amortized model alone.

\begin{figure*}[t!]
    \centering
    \includegraphics[width=0.7\linewidth]{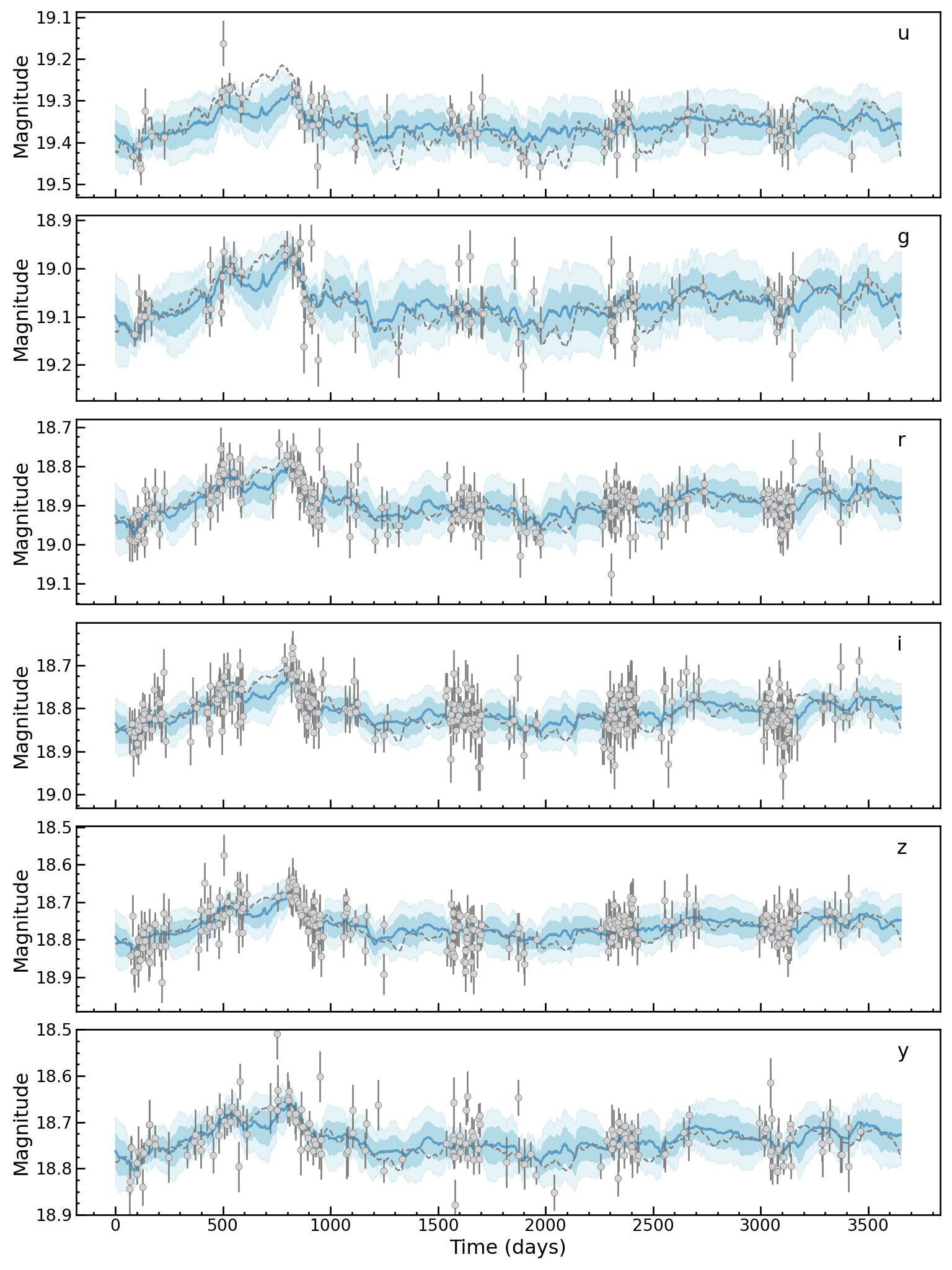}
    \caption{
Example reconstruction of a simulated multi-band LSST AGN light curve from the testing set. Each panel shows one LSST photometric band ($u$ through $y$) as a function of time. Gray points with error bars represent the simulated LSST observations, and dashed gray curves show the underlying simulated light curves prior to observational sampling. Solid blue curves show the ANP posterior mean prediction, with shaded blue regions indicating the $1\sigma$ and $2\sigma$(lighter blue) predictive uncertainty. The ANP reconstruction closely follows the underlying light curves across all bands, demonstrating that the model can recover the temporal variability structure from irregularly sampled, noisy multi-band observations. The predictive uncertainty is reduced near epochs with observations, consistent with expected posterior contraction when the model is conditioned on nearby measurements.}
    \label{fig:LC_reconstruction}
\end{figure*}

\begin{figure}[t!]
    \centering
    \includegraphics[width=1.0\linewidth]{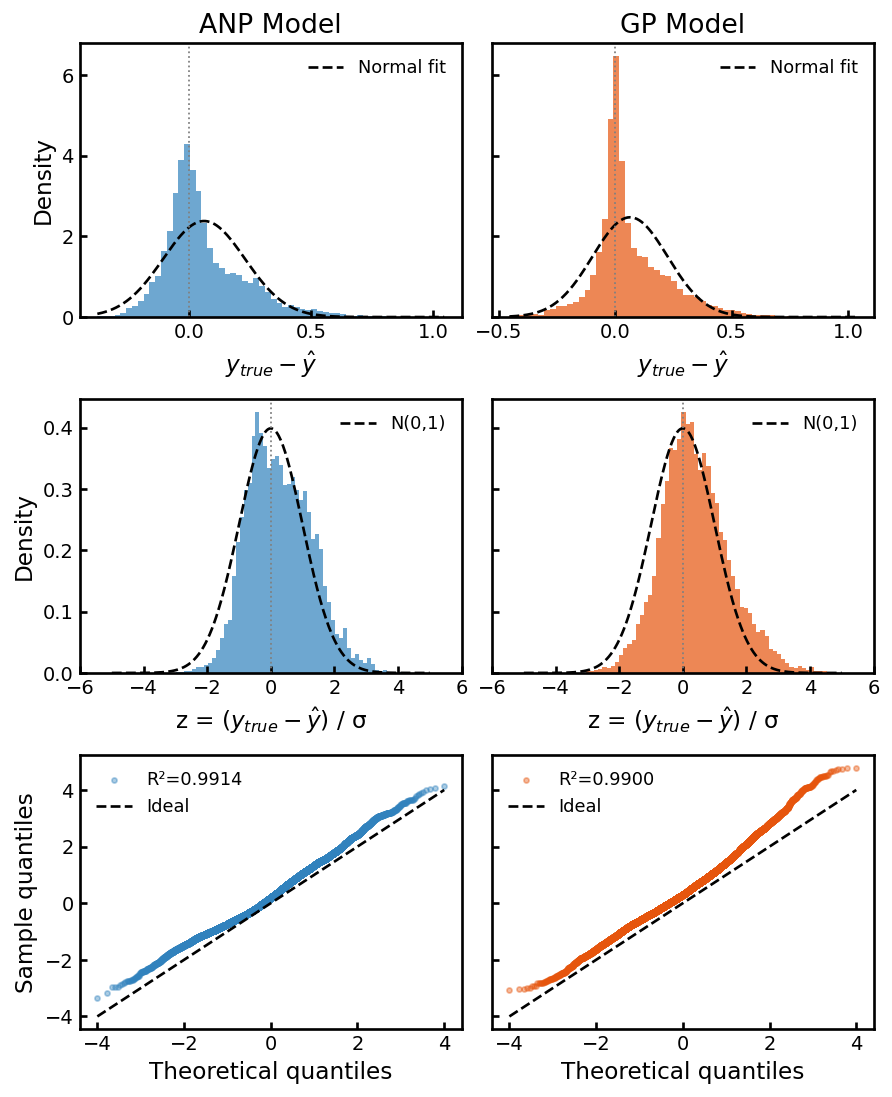}
    \caption{
Residual analysis for the ANP model (left column) and the baseline GP model (right column). The top row shows histograms of raw residuals overlaid with fitted normal distributions (black dashed curves). The middle row shows standardized residual histograms overlaid with the standard normal $\mathcal{N}(0,1)$. The bottom row shows normal Q--Q plots of the standardized residuals against theoretical standard normal quantiles; points lying along the diagonal (dashed line) indicate perfect Gaussian calibration. The ANP model achieves
a Q--Q $R^2 = 0.9914$ compared to $0.9900$ for the GP, confirming better uncertainty calibration in this particular object.}
    \label{fig:residual_analysis}
\end{figure}

\begin{figure}[t!]
    \centering
    \includegraphics[width=1.0\linewidth]{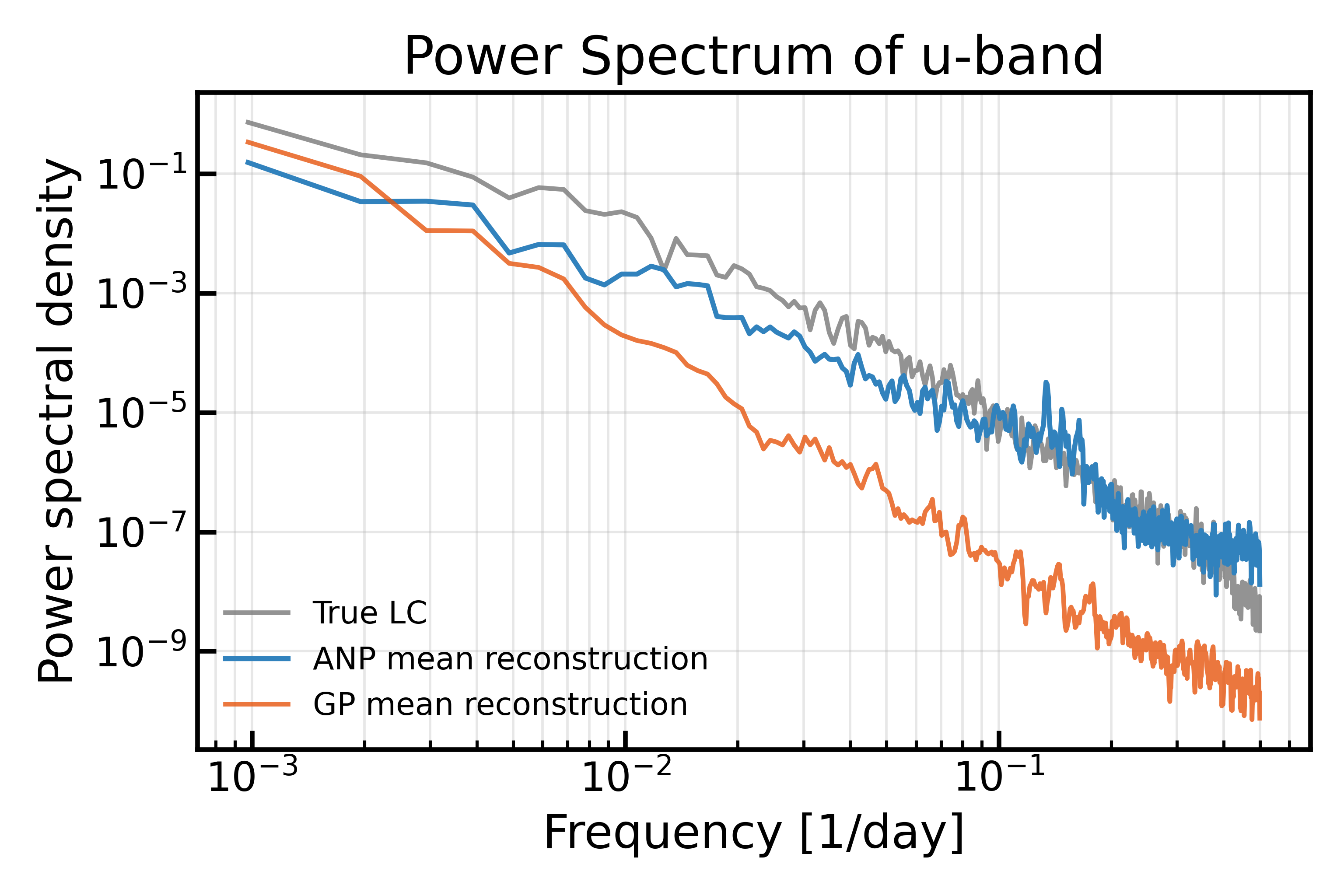}
    \caption{
Power spectral density of the $u$-band for the true underlying light curve (gray), the ANP reconstruction (blue), and the baseline Gaussian Process (GP) reconstruction (orange). The power spectrum is calculated based on mean reconstruction using Welch method with frequency bin size of $1/1024 ~day^{-1}$. The ANP model more faithfully reproduces the power spectrum of the true light curve at high frequencies, indicating better recovery of short-timescale variability.}
    \label{fig:Power_spectrum}
\end{figure}

\begin{table}[hb]
\centering
\begin{tabular}{lcc}
\hline
\textbf{Metric} & \textbf{ANP} & \textbf{GP} \\
\hline
MAE           & 0.122  & 0.115 \\
RMSE          & 0.178  & 0.174 \\
Mean residual & 0.0605 & 0.0637 \\
Std($z$)      & 1.01   & 1.08 \\
Mean $|z|$    & 0.837  & 0.880 \\
\hline
\end{tabular}
\caption{Summary of error and calibration metrics for the ANP and baseline GP models. Both models exhibit small mean residuals, indicating negligible systematic bias. While the GP achieves lower MAE and RMSE, the ANP model produces standardized residuals closer to a standard normal distribution ($\mathrm{Std}(z) \approx 1$, $\mathrm{Mean}\,|z| \approx 0.8$), indicating better-calibrated uncertainty estimates.}
\label{tab:error_analysis}
\end{table}

\subsection{Light-curve Reconstruction}
We next evaluate the ability of the attentive neural process model to reconstruct LSST-like AGN light curves.
% \cx{Comment on the effect of the high frequency end of damped random walk model, and also the baseline limitation.}  

Fig.~\ref{fig:LC_reconstruction} shows an example reconstruction for a light curve from the testing set. Across all six LSST bands, the calibrated posterior mean predicted by the model closely follows the underlying simulated light curves, indicating that the model can recover the temporal variability structure from irregularly sampled and noisy observations. We also find that the predictive uncertainty behaves consistently with the available observational information. In contrast to the model of \cite{park2021inferringblackholeproperties}, which predicts a relatively uniform and broad uncertainty, our model  produces uncertainties that decrease near epochs 
with observed data by introducing the distance feature into MLP (Eq.~\ref{dist_sig}). This local reduction in predictive uncertainty reflects the expected posterior uncertainty shrinkage when the model is conditioned on nearby measurements. Similarly, the uncertainty increases away from observed epochs.

We compare the ANP model against a baseline Gaussian process (GP) across
several metrics. For the GP we adopt the DRW kernel and fit each object and
band individually by maximum a posteriori estimation. Since the simulated
driving signals are themselves DRWs, this baseline is a nearly optimal reference for light curve reconstruction.

Table~\ref{tab:error_analysis} shows point-wise prediction accuracy for the
light curve shown in Fig.~\ref{fig:LC_reconstruction}. The two models
perform comparably
(RMSE $0.174$ versus $0.178$). For this example the ANP produces a mean residual of
$0.061$ and standardized residuals with $\mathrm{Std}(z) = 1.01$ and
$\mathrm{Mean}\,|z| = 0.837$, close to the values of $1$ and
$\sqrt{2/\pi} \approx 0.798$ expected under $\mathcal{N}(0,1)$, indicating
essentially unbiased and well-scaled uncertainties for this object.

Figure~\ref{fig:residual_analysis} extends the residual analysis. Both
models produce similar residual and standardized-residual distributions,
consistent with near-Gaussian behavior. In the normal Q--Q plots the ANP
attains $R^2 = 0.9914$ against $R^2 = 0.9900$ for the GP, implying that both are Gaussian-like.

We next examine the power spectral density of the $u$-band reconstructions
(Fig.~\ref{fig:Power_spectrum}), estimated with the Welch method\citep{welch1967use}, to assess
how much of the temporal variability structure each reconstruction
preserves. In general, the power spectra lose power at frequencies unconstrained by the data; the
relevant comparison is how much constrained power each model retains. The
ANP reconstruction traces the true spectrum substantially further into the
high-frequency regime, whereas the GP reconstruction exhibits suppressed
power beginning at intermediate frequencies
(${\sim}10^{-2}\,\mathrm{day}^{-1}$).

We also report the NLL comparison between our ANP model and the GP model across the entire testing data set in Tab.~\ref{tab:nll_comparison}. We see that overall our ANP model's light curve reconstruction performance can approach per-object DRW Gaussian-process fits — the near-optimal baseline on DRW data.

\begin{figure}[t]
    \centering
    \includegraphics[width=1.0\linewidth]{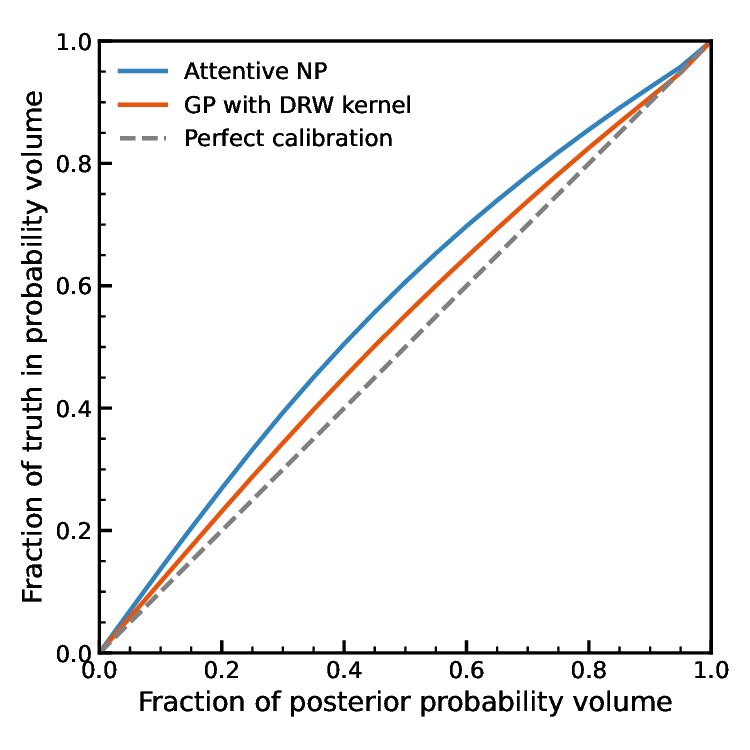}
    \caption{
Calibration curve for light curve reconstruction over the full validation set, 
comparing the ANP (blue) and the baseline GP model (orange) against perfect calibration (gray dashed diagonal). The $x$-axis shows the nominal posterior probability volume $p$, and the $y$-axis shows the empirical fraction of true underlying light curve values falling within the central $p$ credible interval of the predictive posterior. A perfectly calibrated model should lie along the diagonal, where the stated credible intervals contain the true values at exactly the corresponding rate.
} 
    \label{fig:LC_recon_calibration}
\end{figure}

Finally, Fig.~\ref{fig:LC_recon_calibration} assesses the calibration of
the predictive uncertainties over the full testing set, evaluated against
the true underlying light curves on a dense time grid spanning observed
and unobserved epochs. For a nominal credible level $p$, the curve shows
the empirical fraction of true values contained within the central $p$
credible interval. Both models lie on or above the diagonal at all
credible levels --- neither is over-confident anywhere --- with the GP,
as expected for the matched kernel, closest to perfect calibration, and
the ANP somewhat more conservative: its 68\% (95\%) intervals contain
$76.38\%$($95.77\%$) of the true values, versus
$72.05\%$ ($94.82\%$) for the GP.

The Q--Q plots,
standardized-residual statistics, the NLL calculation and calibration curves show that the
ANP matches the near-oracle GP in reconstruction accuracy and residual
Gaussianity while providing modestly conservative, never over-confident,
uncertainty estimates across epochs, bands, and objects.

\section{Discussion and Future Work}
\label{sec:discussion}

In this work, we present an upgraded attentive neural process (ANP) model
based on the framework of \citet{park2021inferringblackholeproperties}. The
model is designed to reconstruct AGN light curves across the six LSST bands
and to infer physical and variability parameters of the central black hole
system. Compared with the original ANP framework, our model is trained on
more realistic simulated LSST light curves that include wavelength-dependent
transfer functions, the fielded LSST observing cadence, and per-visit
photometric noise. Most importantly, our data representation explicitly
accounts for the asynchronous sampling of different photometric bands, which
is a key feature of realistic LSST observations.

The light-curve reconstruction component of the model is trained in an
self-supervised manner, using realistic simulated observations that include
LSST-like cadence and noise properties, rather than the noiseless continuous
light curves themselves. As a result, the model learns to infer the latent
temporal structure of the source solely from sparse, noisy, and irregularly
sampled light curves. In contrast, the parameter-inference decoder is
trained in a supervised manner using the known simulation parameters. This
hybrid training strategy allows the model to pre-learn a compact latent
representation that supports both light-curve reconstruction and Bayesian
parameter inference. In future work, this framework can be further trained
or fine-tuned using the millions of real AGN light curves expected from
LSST, as well as the smaller subset of objects with independently inferred
physical properties, for example from spectroscopic measurements.

Our work belongs to a growing family of neural-process approaches to AGN
variability. \citet{cvorovic2021conditional} first applied conditional
neural processes to non-parametric AGN light-curve modeling, demonstrating
flexible interpolation of survey light curves, though without parameter
inference or an assessment of predictive-uncertainty calibration.
\citet{raju2026metalearning} recently combined attentive latent neural
processes with self-organizing-map clustering and mixture density networks
to recover accretion-disk transfer functions and SMBH parameters, reporting
distribution-level agreement between predicted and true parameter
populations. Our approach is complementary along several axes: we model the
six bands jointly within a single attention mechanism rather than per band,
train on visit lists drawn from the fielded Rubin observing strategy with
per-visit noise rather than on synthetic burst cadences with noiseless
targets, and evaluate per-object posterior quality --- normalized RMSE and
empirical coverage of individual credible intervals --- which is a stricter
criterion than population-level distributional agreement. Conversely,
\citet{raju2026metalearning} recover the transfer function itself, which our
model treats as part of the simulator, and their clustering-based framework
is designed to operate in the few-hundred-light-curve regime, whereas ours
assumes abundant training simulations.

Compared with the latent-SDE framework of \citet{Fagin_2025}, our ANP model
provides a complementary approach that emphasizes architectural compactness,
efficient amortized inference, and reusable latent representations. The
latent-SDE model embeds a physically motivated continuous-time stochastic
process and an auto-differentiable accretion-disk reprocessing model within
the neural-network architecture, providing a highly self-consistent physical
framework. In contrast, our model uses the physical reprocessing model to
generate realistic training data, but does not require solving a latent SDE
or evaluating transfer-function convolutions during inference. As a result,
the ANP model has fewer trainable parameters and is faster to train in our
implementation: while the latent-SDE model was reported to require
approximately three weeks of training \citep{Fagin_2025}, our ANP model
reaches convergence in approximately thirty hours, providing a
computationally efficient pathway for future fine-tuning on real LSST data.
The parameter-inference and reconstruction performance we obtain is
qualitatively comparable to that reported by \citet{Fagin_2025}.

The compact latent representation learned by the ANP also provides a useful
summary of the information contained in the light curves. As demonstrated by
the black hole parameter-inference results, this latent representation
encodes information about the host AGN system and its variability. This
makes the representation useful not only for light-curve reconstruction, but
also for additional downstream tasks. In future work, we plan to explore
applications such as anomaly detection, binary AGN classification, and
identifying unusual variability patterns in large-scale time-domain surveys.

Overall, our results suggest that the upgraded ANP model provides a flexible
and efficient framework for simulation-based AGN light-curve analysis. Its
ability to handle irregular, asynchronous multi-band observations makes it
particularly well suited to LSST-like data. A natural next step is to apply
the model to real AGN light curves and to investigate simulation-to-real
transfer, including fine-tuning strategies in which model selection relies
only on observable quantities, as no noiseless ground truth exists for real
data. This provides a pathway toward using simulation-trained neural-process
models for large-scale inference on real time-domain survey data.
\begin{table}[t]
    \centering
    \label{tab:nll_comparison}
    \begin{tabular}{lcc}
        \toprule
        Metric & ANP & GP \\
        \midrule
        Mean NLL & $-2.35$ & $-2.36$ \\
        Std NLL  & $ 0.983$  & $0.984$ \\
        Median NLL & $-2.47$ & $-2.40$\\
        Meadian absolute deviation & $0.810$ &$ 0.835$ \\
        Min NLL  & $-3.95$ & $-4.10$ \\
        Max NLL  & $-0.315$  & $-0.427$ \\
        \bottomrule
    \end{tabular}
    \caption{Negative Log-Likelihood (NLL) per Light curve object(Averaged over time and bands) for the ANP and baseline GP models evaluated on the whole testing set. Lower NLL indicates better probabilistic predictive performance.}
\end{table}

% \begin{itemize}
%     \item compare our findings with macleod 2010 (DRW) and Yu 2022 (DHO) 
%     \item application to binary periodicity measurement 
%     \item difference in methodology and key findings between this work and Fagin 25 + Park 21. I suggest shortening the discussions in the introduction and move most of the discussions here. \cx{comment on the efficiency difference and accuracy for parameter inference. }
%     \item applications to other surveys (ZTF)
% \end{itemize}

%% For this sample we use BibTeX plus aasjournalv7.bst to generate the

\begin{acknowledgments}

This work made use of Claude Code (Anthropic) for assistance with code
development, data-validation tooling, and manuscript polishing. All results
were verified by the authors, who take full responsibility for the content
of this paper.

The Villar Astro Time Lab acknowledges support through the David and Lucile Packard Foundation, the Research Corporation for Scientific Advancement (through a Cottrell Fellowship and Bridge Award) and the National Science Foundation under AST-2433718, AST-2407922 and AST-2406110. This work is supported
by the National Science Foundation under Cooperative Agreement PHY-2019786 (the NSF AI Institute for
Artificial Intelligence and Fundamental Interactions). T.S. Thanks the Ph.D. fellowship support from Think Global Education Trust. C.X. thanks the support by the Black Hole Initiative, which is funded by grants from the John Templeton Foundation (Grant \#63445) and the Gordon and Betty Moore Foundation
(Grant \#13526), although the opinions expressed
in this work are those of the author(s) and do not necessarily reflect the views of these Foundations. KESF is supported by NSF AST-2206096, NSF AST-1831415 and Simons Foundation Grant 533845. Support was provided by Schmidt Sciences, LLC. for J.F. 
\end{acknowledgments} 

\newpage
\bibliography{sample701}{}

@article{yu2025examining,
  title={Examining Active Galactic Nucleus UV/Optical Variability beyond the Simple Damped Random Walk. II. Insights from 22 yr Observations of SDSS, PS1, and ZTF},
  author={Yu, Weixiang and Richards, Gordon T and Ruan, John J and Vogeley, Michael S and Bauer, Franz E and Graham, Matthew J},
  journal={The Astrophysical Journal},
  volume={992},
  number={1},
  pages={130},
  year={2025},
  publisher={The American Astronomical Society}
}

@article{Fagin_2025,
doi = {10.3847/1538-4357/addabc},
url = {https://doi.org/10.3847/1538-4357/addabc},
year = {2025},
month = {jul},
publisher = {The American Astronomical Society},
volume = {988},
number = {1},
pages = {59},
author = {Fagin, Joshua and Chan, James Hung-Hsu and Best, Henry and O’Dowd, Matthew and Ford, K. E. Saavik and Graham, Matthew J. and Park, Ji Won and Villar, V. Ashley},
title = {Joint Modeling of Quasar Variability and Accretion Disk Reprocessing Using Latent Stochastic Differential Equations},
journal = {The Astrophysical Journal}
}

@article{Chan_2025,
doi = {10.3847/1538-4357/ad93af},
url = {https://doi.org/10.3847/1538-4357/ad93af},
year = {2024},
month = {dec},
publisher = {The American Astronomical Society},
volume = {978},
number = {1},
pages = {54},
author = {Chan, James Hung-Hsu and Fagin, Joshua and Best, Henry and O’Dowd, Matthew J.},
title = {Reverberation Mapping of Lamppost and Wind Structures in Accretion Thin Disks},
journal = {The Astrophysical Journal}
}

@misc{park2021inferringblackholeproperties,
      title={Inferring Black Hole Properties from Astronomical Multivariate Time Series with Bayesian Attentive Neural Processes}, 
      author={Ji Won Park and Ashley Villar and Yin Li and Yan-Fei Jiang and Shirley Ho and Joshua Yao-Yu Lin and Philip J. Marshall and Aaron Roodman},
      year={2021},
      eprint={2106.01450},
      archivePrefix={arXiv},
      primaryClass={astro-ph.IM},
      url={https://arxiv.org/abs/2106.01450}, 
}

@misc{dubois2020npf,
  title        = {Neural Process Family},
  author       = {Dubois, Yann and Gordon, Jonathan and Foong, Andrew YK},
  month        = {September},
  year         = {2020},
  howpublished = {\url{http://yanndubs.github.io/Neural-Process-Family/}}
}

@misc{SMTN002,
  author       = {Jones, R. Lynne},
  title        = {{Calculating LSST Limiting Magnitudes and SNR}},
  howpublished = {{Rubin Observatory Simulations Technical Note SMTN-002}},
  year         = {2022},
  url          = {https://smtn-002.lsst.io/},
  note         = {Accessed 2026-07-14}
}

@article{VandenBerk2004,
  author  = {Vanden Berk, Daniel E. and Wilhite, Brian C. and Kron, Richard G.
             and Anderson, Scott F. and Brunner, Robert J. and Hall, Patrick B.
             and Ivezi{\'c}, {\v{Z}}eljko and Richards, Gordon T.
             and Schneider, Donald P. and York, Donald G.
             and Brinkmann, Jonathan V. and Lamb, Don Q.
             and Nichol, Robert C. and Schlegel, David J.},
  title   = {The Ensemble Photometric Variability of Approximately 25,000
             Quasars in the Sloan Digital Sky Survey},
  journal = {The Astrophysical Journal},
  year    = {2004},
  volume  = {601},
  number  = {2},
  pages   = {692--714},
  doi     = {10.1086/380563},
  eprint  = {astro-ph/0310336},
  archiveprefix = {arXiv}
}

@article{Wilhite2005,
  author  = {Wilhite, Brian C. and Vanden Berk, Daniel E.
             and Kron, Richard G. and Schneider, Donald P.
             and Pereyra, Nicholas and Brunner, Robert J.
             and Richards, Gordon T. and Brinkmann, Jonathan V.},
  title   = {Spectral Variability of Quasars in the Sloan Digital Sky Survey.
             I. Wavelength Dependence},
  journal = {The Astrophysical Journal},
  year    = {2005},
  volume  = {633},
  number  = {2},
  pages   = {638--648},
  doi     = {10.1086/430821},
  eprint  = {astro-ph/0504309},
  archiveprefix = {arXiv}
}

@article{Soldi2008,
  author  = {Soldi, S. and T{\"u}rler, M. and Paltani, S.
             and Aller, H. D. and Aller, M. F. and Burki, G.
             and Chernyakova, M. and L{\"a}hteenm{\"a}ki, A.
             and McHardy, I. M. and Robson, E. I.
             and Staubert, R. and Tornikoski, M.
             and Walter, R. and Courvoisier, T. J.-L.},
  title   = {The Multiwavelength Variability of {3C 273}},
  journal = {Astronomy \& Astrophysics},
  year    = {2008},
  volume  = {486},
  number  = {2},
  pages   = {411--425},
  doi     = {10.1051/0004-6361:200809947},
  eprint  = {0805.3411},
  archiveprefix = {arXiv},
  primaryclass = {astro-ph}
}

@article{Kelly2009,
  author  = {Kelly, Brandon C. and Bechtold, Jill
             and Siemiginowska, Aneta},
  title   = {Are the Variations in Quasar Optical Flux Driven by Thermal
             Fluctuations?},
  journal = {The Astrophysical Journal},
  year    = {2009},
  volume  = {698},
  number  = {1},
  pages   = {895--910},
  doi     = {10.1088/0004-637X/698/1/895},
  eprint  = {0903.5315},
  archiveprefix = {arXiv},
  primaryclass = {astro-ph.CO}
}

@article{MacLeod2010,
  author  = {MacLeod, Chelsea L. and Ivezi{\'c}, {\v{Z}}eljko
             and Kochanek, Christopher S. and Koz{\l}owski, Szymon
             and Kelly, Brandon and Bullock, Elizabeth
             and Kimball, Amy and Sesar, Branimir
             and Westman, Dan and Brooks, Kelsey
             and Gibson, Robert and Becker, Andrew C.
             and de Vries, W. H.},
  title   = {Modeling the Time Variability of {SDSS} Stripe 82 Quasars
             as a Damped Random Walk},
  journal = {The Astrophysical Journal},
  year    = {2010},
  volume  = {721},
  number  = {2},
  pages   = {1014--1033},
  doi     = {10.1088/0004-637X/721/2/1014},
  eprint  = {1004.0276},
  archiveprefix = {arXiv},
  primaryclass = {astro-ph.CO}
}

@article{Mushotzky2011,
  author  = {Mushotzky, Richard F. and Edelson, Rick
             and Baumgartner, Wayne and Gandhi, Poshak},
  title   = {{Kepler} Observations of Rapid Optical Variability in Active
             Galactic Nuclei},
  journal = {The Astrophysical Journal Letters},
  year    = {2011},
  volume  = {743},
  number  = {1},
  pages   = {L12},
  doi     = {10.1088/2041-8205/743/1/L12},
  eprint  = {1111.0672},
  archiveprefix = {arXiv},
  primaryclass = {astro-ph.CO}
}

@article{Yu2022,
  author  = {Yu, Weixiang and Richards, Gordon T.
             and Vogeley, Michael S. and Moreno, Jackeline
             and Graham, Matthew J.},
  title   = {Examining {AGN} {UV}/Optical Variability beyond the Simple
             Damped Random Walk},
  journal = {The Astrophysical Journal},
  year    = {2022},
  volume  = {936},
  number  = {2},
  pages   = {132},
  doi     = {10.3847/1538-4357/ac8351},
  eprint  = {2201.08943},
  archiveprefix = {arXiv},
  primaryclass = {astro-ph.GA}
}

@article{ShakuraSunyaev1973,
  author  = {Shakura, N. I. and Sunyaev, R. A.},
  title   = {Black Holes in Binary Systems: Observational Appearance},
  journal = {Astronomy \& Astrophysics},
  year    = {1973},
  volume  = {24},
  pages   = {337--355},
  adsurl  = {https://ui.adsabs.harvard.edu/abs/1973A%26A....24..337S}
}

@article{Pereyra2006,
  author        = {Pereyra, Nicolas A. and Vanden Berk, Daniel E.
                   and Turnshek, David A. and Hillier, D. John
                   and Wilhite, Brian C. and Kron, Richard G.
                   and Schneider, Donald P. and Brinkmann, Jonathan},
  title         = {Characteristic {QSO} Accretion Disk Temperatures from
                   Spectroscopic Continuum Variability},
  journal       = {The Astrophysical Journal},
  year          = {2006},
  volume        = {642},
  number        = {1},
  pages         = {87--95},
  doi           = {10.1086/500919},
  eprint        = {astro-ph/0506006},
  archiveprefix = {arXiv}
}

@article{Liu2008,
  author        = {Liu, H. T. and Bai, J. M. and Zhao, X. H. and Ma, L.},
  title         = {Tests for Standard Accretion Disk Models by Variability
                   in Active Galactic Nuclei},
  journal       = {The Astrophysical Journal},
  year          = {2008},
  volume        = {677},
  number        = {2},
  pages         = {884--894},
  doi           = {10.1086/529361},
  eprint        = {0803.0356},
  archiveprefix = {arXiv},
  primaryclass  = {astro-ph}
}

@article{Wanders1997,
  author  = {Wanders, I. and Peterson, B. M. and Alloin, D.
             and Ayres, T. R. and Clavel, J. and Crenshaw, D. M.
             and Horne, K. and Kriss, G. A. and Krolik, J. H.
             and Malkan, M. A. and Netzer, H. and O'Brien, P. T.
             and Reichert, G. A. and Rodr{\'i}guez-Pascual, P. M.
             and Wamsteker, W. and others},
  title   = {Steps toward Determination of the Size and Structure of the
             Broad-Line Region in Active Galactic Nuclei. XI. Intensive
             Monitoring of the Ultraviolet Spectrum of {NGC 7469}},
  journal = {The Astrophysical Journal Supplement Series},
  year    = {1997},
  volume  = {113},
  number  = {1},
  pages   = {69--88},
  doi     = {10.1086/313054}
}

@article{BlandfordMcKee1982,
  author  = {Blandford, Roger D. and McKee, Christopher F.},
  title   = {Reverberation Mapping of the Emission Line Regions of
             {Seyfert} Galaxies and Quasars},
  journal = {The Astrophysical Journal},
  year    = {1982},
  volume  = {255},
  pages   = {419--439},
  doi     = {10.1086/159843}
}

@article{Cackett2021,
  author        = {Cackett, Edward M. and Bentz, Misty C. and Kara, Erin},
  title         = {Reverberation Mapping of Active Galactic Nuclei:
                   From {X}-Ray Corona to Dusty Torus},
  journal       = {iScience},
  year          = {2021},
  volume        = {24},
  number        = {6},
  pages         = {102557},
  doi           = {10.1016/j.isci.2021.102557},
  eprint        = {2105.06926},
  archiveprefix = {arXiv},
  primaryclass  = {astro-ph.GA}
}

@article{Wang2023,
  author        = {Wang, Shu and Guo, Hengxiao and Woo, Jong-Hak},
  title         = {Estimating {AGN} Black Hole Masses via Continuum
                   Reverberation Mapping in the Era of {LSST}},
  journal       = {The Astrophysical Journal Letters},
  year          = {2023},
  volume        = {948},
  number        = {2},
  pages         = {L23},
  doi           = {10.3847/2041-8213/accf96},
  eprint        = {2302.05261},
  archiveprefix = {arXiv},
  primaryclass  = {astro-ph.GA}
}

@article{LyndenBell1969,
  author  = {Lynden-Bell, Donald},
  title   = {Galactic Nuclei as Collapsed Old Quasars},
  journal = {Nature},
  year    = {1969},
  volume  = {223},
  pages   = {690--694},
  doi     = {10.1038/223690a0}
}

@article{Netzer2015,
  author  = {Netzer, Hagai},
  title   = {Revisiting the Unified Model of Active Galactic Nuclei},
  journal = {Annual Review of Astronomy and Astrophysics},
  year    = {2015},
  volume  = {53},
  pages   = {365--408},
  doi     = {10.1146/annurev-astro-082214-122302},
  eprint  = {1505.00811},
  archiveprefix = {arXiv},
  primaryclass  = {astro-ph.GA}
}

@article{Ulrich1997,
  author  = {Ulrich, Marie-Helene and Maraschi, Laura and Urry, C. Megan},
  title   = {Variability of Active Galactic Nuclei},
  journal = {Annual Review of Astronomy and Astrophysics},
  year    = {1997},
  volume  = {35},
  pages   = {445--502},
  doi     = {10.1146/annurev.astro.35.1.445}
}

@article{Giveon1999,
  author  = {Giveon, U. and Maoz, D. and Kaspi, S.
             and Netzer, H. and Smith, P. S.},
  title   = {Long-Term Optical Variability Properties of the
             {Palomar-Green} Quasars},
  journal = {Monthly Notices of the Royal Astronomical Society},
  year    = {1999},
  volume  = {306},
  number  = {3},
  pages   = {637--654},
  doi     = {10.1046/j.1365-8711.1999.02556.x},
  eprint  = {astro-ph/9902254},
  archiveprefix = {arXiv}
}

@article{McHardy2006,
  author  = {McHardy, I. M. and Koerding, E.
             and Knigge, C. and Uttley, P. and Fender, R. P.},
  title   = {Active Galactic Nuclei as Scaled-Up Galactic Black Holes},
  journal = {Nature},
  year    = {2006},
  volume  = {444},
  number  = {7120},
  pages   = {730--732},
  doi     = {10.1038/nature05389},
  eprint  = {astro-ph/0612273},
  archiveprefix = {arXiv}
}

@article{Edelson2015,
  author  = {Edelson, R. and Gelbord, J. M. and Horne, K.
             and McHardy, I. M. and Peterson, B. M.
             and Ar{\'e}valo, P. and others},
  title   = {Space Telescope and Optical Reverberation Mapping Project.
             II. {Swift} and {HST} Reverberation Mapping of the
             Accretion Disk of {NGC 5548}},
  journal = {The Astrophysical Journal},
  year    = {2015},
  volume  = {806},
  number  = {1},
  pages   = {129},
  doi     = {10.1088/0004-637X/806/1/129},
  eprint  = {1501.05951},
  archiveprefix = {arXiv},
  primaryclass  = {astro-ph.GA}
}

@article{Xin2021,
archivePrefix = {arXiv},
arxivId = {2105.00005},
author = {Xin, Chengcheng and Haiman, Zolt{\'{a}}n},
doi = {10.1093/mnras/stab1856},
eprint = {2105.00005},
issn = {13652966},
journal = {Monthly Notices of the Royal Astronomical Society},
number = {2},
pages = {2408--2417},
title = {{Ultra-short-period massive black hole binary candidates in LSST as LISA 'verification binaries'}},
volume = {506},
year = {2021}
}

@article{Viitanen2026AGILE,
  author = {Viitanen, A. and Bongiorno, A. and Saccheo, I. and Grazian, A.
            and Paolillo, M. and Petrecca, V. and De Cicco, D. and Roberts, D.
            and Shankar, F. and Allevato, V. and Merlin, E. and Ili{\'c}, D.
            and Kova{\v{c}}evi{\'c}, A. B. and De Somma, G.
            and Di Criscienzo, M. and Girardi, L. and Marconi, M. and Mazzi, A.
            and Pastorelli, G. and Trabucchi, M. and Ananna, T. and Assef, R. J.
            and Brandt, W. N. and Brescia, M. and Graham, A. W. and Li, G.
            and Marsango, D. and Peca, A. and Polioudakis, M.
            and Raiteri, C. M. and Rani, B. and Ricci, C. and Richards, G.
            and Salvato, M. and Satheesh-Sheeba, S. and Shirley, R. and Tang, S.
            and Temple, M. J. and Tombesi, F. and Yoon, I. and Zou, F.},
  title = {{AGILE: An end-to-end Rubin-LSST simulation of active galactic nuclei,
            galaxies, and stars. I. Software description and first data release}},
  journal = {Astronomy \& Astrophysics},
  year = {2026},
  volume = {710},
  pages = {A56},
  doi = {10.1051/0004-6361/202558627}
}

@article{LSSTScienceCollaboration2009,
archivePrefix = {arXiv},
arxivId = {0912.0201},
author={{LSST Science Collaboration} and Abell, Paul A. and Allison, Julius and Anderson, Scott F. and Andrew, John R. and Angel, J. Roger P. and Armus, Lee and Arnett, David and Asztalos, S. J. and Axelrod, Tim S. and Bailey, Stephen and Ballantyne, D. R. and Bankert, Justin R. and Barkhouse, Wayne A. and Barr, Jeffrey D. and Barrientos, L. Felipe and Barth, Aaron J. and Bartlett, James G. and Becker, Andrew C. and Becla, Jacek and Beers, Timothy C. and Bernstein, Joseph P. and Biswas, Rahul and Blanton, Michael R. and Bloom, Joshua S. and Bochanski, John J. and Boeshaar, Pat and Borne, Kirk D. and Bradac, Marusa and Brandt, W. N. and Bridge, Carrie R. and Brown, Michael E. and Brunner, Robert J. and Bullock, James S. and Burgasser, Adam J. and Burge, James H. and Burke, David L. and Cargile, Phillip A. and Chandrasekharan, Srinivasan and Chartas, George and Chesley, Steven R. and Chu, You-Hua and Cinabro, David and Claire, Mark W. and Claver, Charles F. and Clowe, Douglas and Connolly, A. J. and Cook, Kem H. and Cooke, Jeff and Cooray, Asantha and Covey, Kevin R. and Culliton, Christopher S. and de Jong, Roelof and de Vries, Willem H. and Debattista, Victor P. and Delgado, Francisco and Dell'Antonio, Ian P. and Dhital, Saurav and {Di Stefano}, Rosanne and Dickinson, Mark and Dilday, Benjamin and Djorgovski, S. G. and Dobler, Gregory and Donalek, Ciro and Dubois-Felsmann, Gregory and Durech, Josef and Eliasdottir, Ardis and Eracleous, Michael and Eyer, Laurent and Falco, Emilio E. and Fan, Xiaohui and Fassnacht, Christopher D. and Ferguson, Harry C. and Fernandez, Yanga R. and Fields, Brian D. and Finkbeiner, Douglas and Figueroa, Eduardo E. and Fox, Derek B. and Francke, Harold and Frank, James S. and Frieman, Josh and Fromenteau, Sebastien and Furqan, Muhammad and Galaz, Gaspar and Gal-Yam, A. and Garnavich, Peter and Gawiser, Eric and Geary, John and Gee, Perry and Gibson, Robert R. and Gilmore, Kirk and Grace, Emily A. and Green, Richard F. and Gressler, William J. and Grillmair, Carl J. and Habib, Salman and Haggerty, J. S. and Hamuy, Mario and Harris, Alan W. and Hawley, Suzanne L. and Heavens, Alan F. and Hebb, Leslie and Henry, Todd J. and Hileman, Edward and Hilton, Eric J. and Hoadley, Keri and Holberg, J. B. and Holman, Matt J. and Howell, Steve B. and Infante, Leopoldo and Ivezic, Zeljko and Jacoby, Suzanne H. and Jain, Bhuvnesh and R and Jedicke and Jee, M. James and Jernigan, J. Garrett and Jha, Saurabh W. and Johnston, Kathryn V. and Jones, R. Lynne and Juric, Mario and Kaasalainen, Mikko and Styliani and Kafka and Kahn, Steven M. and Kaib, Nathan A. and Kalirai, Jason and Kantor, Jeff and Kasliwal, Mansi M. and Keeton, Charles R. and Kessler, Richard and Knezevic, Zoran and Kowalski, Adam and Krabbendam, Victor L. and Krughoff, K. Simon and Kulkarni, Shrinivas and Kuhlman, Stephen and Lacy, Mark and Lepine, Sebastien and Liang, Ming and Lien, Amy and Lira, Paulina and Long, Knox S. and Lorenz, Suzanne and Lotz, Jennifer M. and Lupton, R. H. and Lutz, Julie and Macri, Lucas M. and Mahabal, Ashish A. and Mandelbaum, Rachel and Marshall, Phil and May, Morgan and McGehee, Peregrine M. and Meadows, Brian T. and Meert, Alan and Milani, Andrea and Miller, Christopher J. and Miller, Michelle and Mills, David and Minniti, Dante and Monet, David and Mukadam, Anjum S. and Nakar, Ehud and Neill, Douglas R. and Newman, Jeffrey A. and Nikolaev, Sergei and Nordby, Martin and O'Connor, Paul and Oguri, Masamune and Oliver, John and Olivier, Scot S. and Olsen, Julia K. and Olsen, Knut and Olszewski, Edward W. and Oluseyi, Hakeem and Padilla, Nelson D. and Parker, Alex and Pepper, Joshua and Peterson, John R. and Petry, Catherine and Pinto, Philip A. and Pizagno, James L. and Popescu, Bogdan and Prsa, Andrej and Radcka, Veljko and Raddick, M. Jordan and Rasmussen, Andrew and Rau, Arne and Rho, Jeonghee and Rhoads, James E. and Richards, Gordon T. and Ridgway, Stephen T. and Robertson, Brant E. and Roskar, Rok and Saha, Abhijit and Sarajedini, Ata and Scannapieco, Evan and Schalk, Terry and Schindler, Rafe and Schmidt, Samuel and Schmidt, Sarah and Schneider, Donald P. and Schumacher, German and Scranton, Ryan and Sebag, Jacques and Seppala, Lynn G. and Shemmer, Ohad and Simon, Joshua D. and Sivertz, M. and Smith, Howard A. and Smith, J. Allyn and Smith, Nathan and Spitz, Anna H. and Stanford, Adam and Stassun, Keivan G. and Strader, Jay and Strauss, Michael A. and Stubbs, Christopher W. and Sweeney, Donald W. and Szalay, Alex and Szkody, Paula and Takada, Masahiro and Thorman, Paul and Trilling, David E. and Trimble, Virginia and Tyson, Anthony and {Van Berg}, Richard and Berk, Daniel Vanden and VanderPlas, Jake and Verde, Licia and Vrsnak, Bojan and Walkowicz, Lucianne M. and Wandelt, Benjamin D. and Wang, Sheng and Wang, Yun and Warner, Michael and Wechsler, Risa H. and West, Andrew A. and Wiecha, Oliver and Williams, Benjamin F. and Willman, Beth and Wittman, David and Wolff, Sidney C. and Wood-Vasey, W. Michael and Wozniak, Przemek and Young, Patrick and Zentner, Andrew and Zhan, Hu},
eprint = {0912.0201},
number = {November},
Journal = {e-print arXiv:0912.0201},
title = {LSST Science Book, Version 2.0},
url = {http://arxiv.org/abs/0912.0201},
year = {2009}
}

@misc{kingma2017improvingvariationalinferenceinverse,
      title={Improving Variational Inference with Inverse Autoregressive Flow}, 
      author={Diederik P. Kingma and Tim Salimans and Rafal Jozefowicz and Xi Chen and Ilya Sutskever and Max Welling},
      year={2017},
      eprint={1606.04934},
      archivePrefix={arXiv},
      primaryClass={cs.LG},
      url={https://arxiv.org/abs/1606.04934}, 
}

@inproceedings{
higgins2017betavae,
title={beta-{VAE}: Learning Basic Visual Concepts with a Constrained Variational Framework},
author={Irina Higgins and Loic Matthey and Arka Pal and Christopher Burgess and Xavier Glorot and Matthew Botvinick and Shakir Mohamed and Alexander Lerchner},
booktitle={International Conference on Learning Representations},
year={2017},
url={https://openreview.net/forum?id=Sy2fzU9gl}
}

@article{welch1967use,
  author  = {Welch, Peter D.},
  title   = {The use of fast {F}ourier transform for the estimation of power
             spectra: A method based on time averaging over short, modified
             periodograms},
  journal = {IEEE Transactions on Audio and Electroacoustics},
  volume  = {15},
  number  = {2},
  pages   = {70--73},
  year    = {1967},
  doi     = {10.1109/TAU.1967.1161901}
}

@article{cvorovic2021conditional,
  author  = {{\v{C}}vorovi{\'c}-Hajdinjak, Iva and Kova{\v{c}}evi{\'c}, An{\dj}elka B. and
             Ili{\'c}, Dragana and Popovi{\'c}, Luka {\v{C}}. and Dai, Xinyu and
             Jankov, Isidora and Radovi{\'c}, Viktor and S{\'a}nchez-S{\'a}ez, Paula and
             Nikutta, Robert},
  title   = {Conditional neural process for nonparametric modeling of active galactic
             nuclei light curves},
  journal = {Astronomische Nachrichten},
  volume  = {343},
  number  = {1-2},
  pages   = {e210103},
  year    = {2022},
  doi     = {10.1002/asna.20210103},
  eprint  = {2111.09751},
  archivePrefix = {arXiv},
  primaryClass  = {astro-ph.GA}
}

@article{raju2026metalearning,
  author  = {Raju, Aman N. and Kova{\v{c}}evi{\'c}, An{\dj}elka B. and Ili{\'c}, Dragana and
             Tombesi, Francesco and Popovi{\'c}, Luka {\v{C}}. and Slezak, Eric and
             Sanchez-Saez, Paula and Pavlovi{\'c}, Marina and
             {\v{C}}vorovi{\'c}-Hajdinjak, Iva and Simi{\'c}, Sa{\v{s}}a and
             Savi{\'c}, {\Dj}or{\dj}e},
  title   = {A Meta-Learning Framework for Multitask Reverberation Mapping in Active
             Galactic Nuclei},
  journal = {arXiv e-prints},
  pages   = {arXiv:2606.09665},
  year    = {2026},
  eprint  = {2606.09665},
  archivePrefix = {arXiv},
  primaryClass  = {astro-ph.GA},
  note    = {submitted to A\&A}
}
\bibliographystyle{aasjournalv7}

\end{document}